\documentclass[letter,12pt]{article}
 \pdfoutput=1
 \usepackage{graphicx}
\usepackage{amsmath,amssymb,verbatim,mathrsfs,cite}
\usepackage{rotating}
\usepackage{multirow}
\usepackage{verbatim}
\def\beq{\begin{equation}}
\def\eeq{\end{equation}}
\def\bea{\begin{eqnarray}}
\def\eea{\end{eqnarray}}
\def\bmat{\begin{pmatrix}}
\def\emat{\end{pmatrix}}
\def\bei{\begin{itemize}}
\def\eei{\end{itemize}}

\def\Emisst{\not  \! \! E_T}

\begin{document}
\baselineskip=18pt
\begin{titlepage}
\noindent

\begin{flushright}
CERN-PH-TH/2011-249\\
MCTP-11-35
\end{flushright}
\vspace{0.5 cm}

\begin{center}
  \begin{Large}
    \begin{bf}
  Theory and LHC Phenomenology of Classicalon Decays
        \end{bf}
  \end{Large}
\end{center}

\begin{center}
\begin{large}

Christophe Grojean$^{a,b}$ and  Rick S. Gupta$^{a,c}$  \\

\end{large}

\vspace{0.3cm}
\begin{it}
$^a$CERN, Theoretical Physics, CH-1211 Geneva 23, Switzerland \\
\vspace{0.1cm}

$^b$Institut de Physique Th\'{e}orique, CEA Saclay, F91191 Gif$-$sur$-$Yvette, France\\
\vspace{0.1cm}
$^c$Michigan Center for Theoretical Physics (MCTP) \\
        ~~University of Michigan, Ann Arbor, MI 48109-1120, USA \\
\end{it}

\vspace{1cm}
\end{center}

\begin{abstract}
\noindent
It has been recently proposed by Dvali et al.~\cite{Dvali:2010jz} that high energy scattering in non-renormalizable theories, like the higgsless Standard Model, can be unitarized by the formation of classical configurations called classicalons. In this work we argue that classicalons should have  analogs of thermodynamic properties like temperature and entropy and perform a model-independent statistical mechanical analysis of classicalon decays. We find that,  in the case of massless quanta, the decay products have a Planck distribution with an effective temperature $T\sim 1/r_*$, where  $r_*$ is the classicalon radius. These results, in particular a computation of the decay multiplicity, $N_*$,  allow us to make the first  collider analysis of classicalization. In the  model for unitarization of $WW$ scattering by classicalization of longitudinal $W$s and $Z$s  we get spectacular multi-$W/Z$ final states that decay into leptons, missing energy  and a very high multiplicity (at least 10) of jets.  We find that  for the classicalization scale, $M_* = v=246$ GeV ($M_*=1$ TeV) discovery should be possible in the present 7 TeV (14 TeV) run of the LHC with  about 10 fb$^{-1}$ (100 fb$^{-1}$) data. We also consider a model to solve the hierarchy problem,  where the classicalons are configurations of the Higgs field which  decay into to multi-Higgs boson final states. We find that,  in this case, for $M_*=500$ GeV ($M_*=1$ TeV), discovery should be possible in the top fusion process with about 10 fb$^{-1}$ (100 fb$^{-1}$) data at 14 TeV LHC.

\end{abstract}

\vspace{3cm}

\begin{flushleft}
\begin{small}
October 2011
\end{small}
\end{flushleft}
\end{titlepage}
\tableofcontents
\newpage

\section{Introduction and motivation}
\label{one}
 To find out how longitudinal $WW$-scattering is unitarized is the \textit{raison d'$\hat{e}$tre} for the LHC.  If the LHC keeps delivering data at the present rate we may know the ultimate fate of the most popular candidate, the Higgs boson, very soon. According to  projections it would be possible to exclude the Standard Model (SM) Higgs boson over the whole mass range with 5 fb$^{-1}$ data although discovery will take some more time~\cite{atlas0}.   An elementary  Higgs boson, however, has its own problems  if it exists as  one must then explain the hierarchy between its mass and the cut-off scale. This suggests the existence of new TeV-scale physics even if the Higgs boson exists.  Thus, whether or not a Higgs exists,  the standard argument goes that that a Wisonian UV completion is required  with new states needing to be integrated in at the TeV scale.  A non-Wilsonian alternative has been proposed in Ref.~\cite{Dvali:2010jz}.  For this the authors take inspiration from the other major problem of high energy physics, that of finding a UV-completion for quantum gravity. It has been argued in Refs.~\cite{Dvali:2010bf,Dvali:2010jz} that in transplanckian 2$\to$2 scattering in gravity there is no violation of perturbative unitarity  because of black hole formation. Black holes are classical objects that decay to many particles and  decays to two particles are suppressed leading to a suppression of the 2$\to$2 scattering amplitudes. As we go to higher energies we get larger black holes and the amplitudes are even more suppressed.  In Refs.~\cite{Dvali:2010jz, Dvali:2010ns, Dvali:2011nj, Dvali:2011th,Bajc:2011ey} it has been proposed that formation of classical objects, called classicalons, is possible in high energy scattering also in non-gravitational theories. This happens if there is a bosonic field (the classicalizer field) which is sourced by derivatively coupled operators that grow with energy. At high enough center of mass energy $\sqrt{\hat{s}}$, the source leads to formation of classical configurations of the classicalizer field.  As the classicalon would in general decay into many particles, the usual problem of perturbative unitarity violation in 2$\to$2 scattering in non-renormalizable theories is thus avoided without a usual Wilsonian UV completion. 
 
 In the case of $WW$-scattering the bosonic field can be  the longitudinal goldstone modes of the $W$. As is well known interactions involving  these modes grow with energy so that an appropriate non-linear interaction can be used for self-sourcing these modes. This way of unitarizing $WW$-scattering  is thus arguably even more economical than having  a single  Higgs. As we will discuss in more detail later, around the classicalization scale the classicalons should be thought of as a tower of quantum resonances and only at energies much higher than this scale do they become truly classical. Thus, whereas around the    classicalization scale, such a theory would resemble standard Wilsonian UV completions, like technicolor, with  resonances appearing at this scale, a  theory  with classicalization would be very different in the deep UV. For instance, the inclusive cross-section in classicalizing theories would grow geometrically as the squared classicalon radius, $r_*^2$, at energies above the classicalization scale, unlike any Wilsonian UV-completion where the cross-section eventually decreases with energy. Classicalization can have an application even if the Higgs boson exists provided appropriate classicalizing interactions are also present. Classical configurations of the Higgs field itself, called Higgsions, can be sourced by the energy of the other SM particles in high energy scattering. The classicalization scale where Higgsion formation starts would then become the scale at which the loop contributions to the Higgs mass get screened,  thus solving the hierarchy problem. The collider signals for these models would be the spectacular production of multiple $W$s and $Z$s in the first case of goldstone classicalization and multi-Higgs final states from Higgsion decays in the second case.

 In this work we want to tackle the important question of classicalon decays. We want to address questions like: How many particles does a classicalon decay to?  What is the energy distribution of these decay products? These questions are important   for understanding both the theory and phenomenology of classicalons. From the theoretical point of view, the most important feature for unitarization of the amplitudes is that a classicalon decays, in general, to many particles and decays to a few particles are suppressed. Thus  understanding  classicalon decays is very important.  From the experimental point of view this is the important ingredient that will allow us to make LHC predictions. This is because while the production cross-section can  be estimated from geometric arguments to be  $\pi r_*^2$, a collider analysis is impossible without knowledge of the multiplicity of the classicalon decay products. We will argue, as was already pointed out in Ref.~\cite{Dvali:2011th},  that classicalons, like black holes, have  properties analogous to entropy and temperature and they decay thermally. This will give us  completely model independent predictions about how classicalons should decay.  Before giving the broad argument that tells us  why classicalons should  have thermodynamic properties we will briefly describe  how classicalization takes place.

We take the  simple example of a massless scalar theory with a single non-linear, non-renormalizable interaction,
\beq
{\cal L} = \partial_\mu \phi \partial^\mu \phi+\frac{ (\partial_\mu \phi \partial^\mu \phi)^2}{M_*^4}.
\eeq
A non-linear 	interaction of a similar form will be used for  classicalization of longitudinal $W$s and $Z$s later. We know that the non-renormalizable term $(\partial_\mu \phi \partial^\mu \phi)^2/M_*^4$ above would become important at length scales smaller than the quantum length cut-off, $L_*=1/M_*$. This  term can  actually become important at even larger length scales, as shown in Refs.~\cite{Dvali:2010jz,Dvali:2010ns}, if $\phi$ takes a large classical value. An analysis in Ref.~\cite{Dvali:2010ns} shows that this is precisely what happens in a scattering process with initial energy bigger than the cut off, i.e. $\sqrt{\hat{s}}>M_*$.\footnote{For recent work on the dynamics of classicalization see Refs.~Ê\cite{Akhoury:2011en,Brouzakis:2011zs}.} The authors solve classical equations of motion to show that if we start with  free spherical wave-packet $\phi_0$ there would be a correction due to the non-linear term,
\beq
\phi=\phi_0+\phi_1
\eeq
that becomes important (i.e $\phi_1 \sim \phi_0$) at a length scale,
\beq
r_*=\frac{\sqrt{\hat{s}}^{~\alpha}}{M_*^{1+\alpha}}
\label{main}
\eeq
where $\alpha$ (always $\leq 1$) is a positive number  that depends on the choice of  non-linear term, and is 1/3 in this example. We can see from the expression above that for $\sqrt{\hat{s}} > M_*$ we get $r_* > L_*=1/M_*$  so that $r_*$ is in fact  a classical length. At distances smaller than $r_*$ the non-linear term becomes important leading to a formation of a classical configuration of radius $r_*$. As is clear from Eq.(\ref{main}) with increasing energy the source due to the non-linear term becomes bigger and bigger in magnitude and  the radius $r_*$ of the classical object increases. This means that with higher energy we do not probe shorter distances in these theories. Black holes are seen as a special case of classicalization  where $r_*$ is the Schwarzschild radius, $M_*$ is $M_{pl}$, the planck mass, and $\alpha=1$. As shown in Ref.~\cite{Dvali:2010jz}  the phenomenon is insensitive to higher order terms in the Lagrangian as these operators give a smaller $r_*$.
\begin{figure}[t]
\begin{center}
\includegraphics[width=0.9\columnwidth]{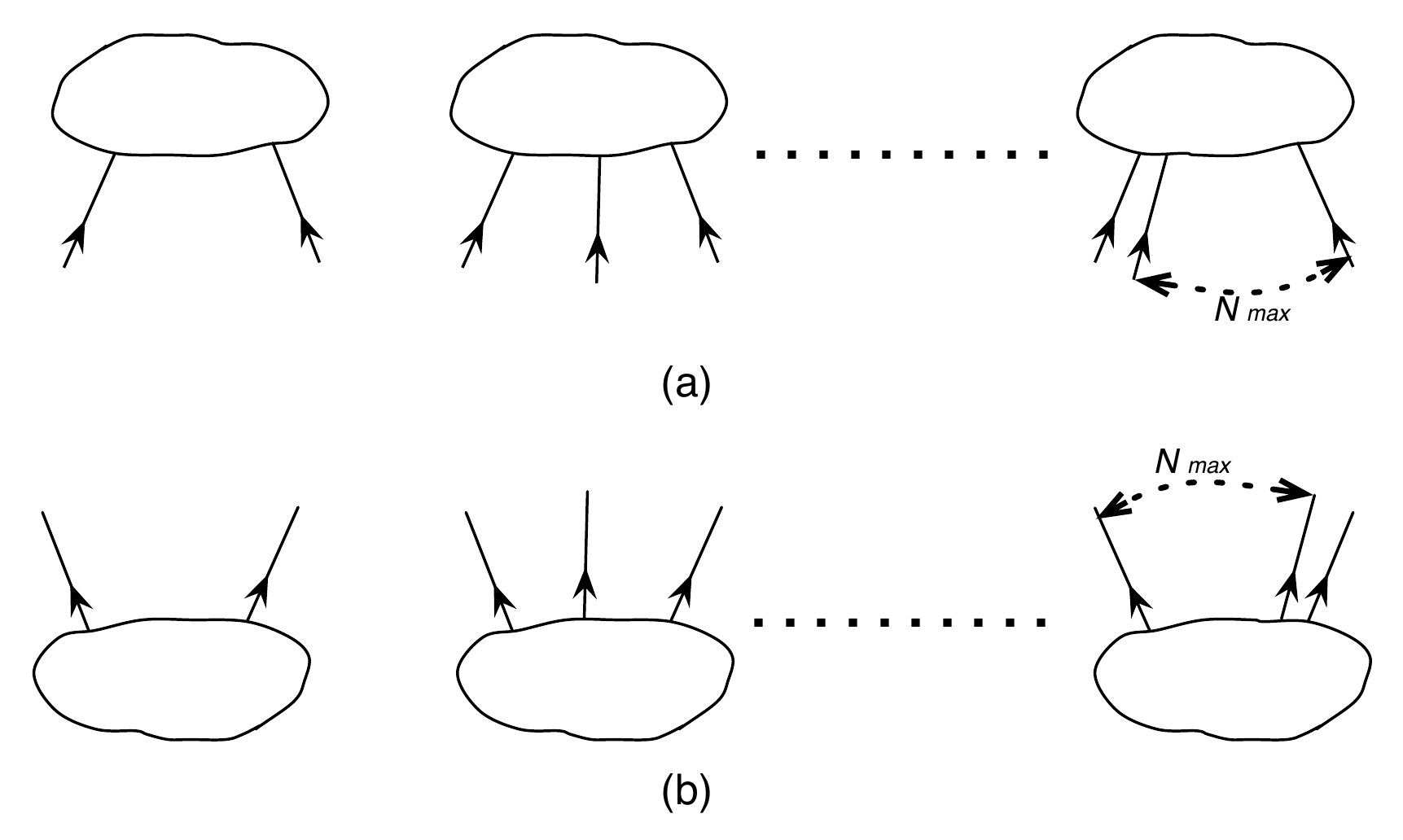}
\end{center}
\caption{(a) Different ways of forming classicalons. Any scattering process with 2,3 ... $N$ initial particles would form a classicalon  if the total energy of these particles. $\sqrt{\hat{s}}$, is larger than   $M_* $. (b) We show the time reverse of the processes shown in Fig. 1 (a). By time reversal symmetry, all these processes should be allowed decays.}
\label{brili}
\end{figure}

We will now motivate why classicalons must have analogs of thermodynamic properties. One way to see how an effective notion of entropy can arise for a classicalon is by noting that there are many ways of forming a classicalon. Any scattering process with 2,3 ... $N$ initial particles shown in Fig.~\ref{brili}(a) would form a classicalon  if the total energy of these particles, $\sqrt{\hat{s}}$, is larger than   $M_* $. There is, however, an upper limit on the number of initial particles. This is because we want the wavelength of the particles $\lambda$ to be smaller than $r_*$, so that the energy of the particles can be localized within the classicalon radius. Assuming massless quanta, the energy of each particle, $1/\lambda$,  must be then at least $1/r_*$. This puts an upper bound on the number $N$ in Fig.~\ref{brili}(a) which is given by,
\beq
N_{max}\sim M/( 1/r_*) \sim M  r_*.
\eeq
where $M$ is the mass of the classicalon. The only restriction on the initial state is the conservation of energy and momentum and ensuring that  the energy of the particles is localized inside the radius $r_*$. We  expect from combinatorics that there would be many more ways of distributing the required energy among many particles than among a few particles, implying that there should be many more ways of forming a classicalon with many particles in the initial state than with a few particles.  Assuming classical time reversal symmetry ($t \to -t$) we can now argue that the time reverse of each of the possible processes shown in Fig.~\ref{brili}(a) is an allowed decay as shown in  Fig.~\ref{brili}(b). Thus it follows that  a classicalon would in general decay to many particles just because of combinatorics. It is also true, however, that just as a classicalon can be formed from two initial particles it can also decay to only two particles but this would be combinatorially suppressed.

 In this work we will find a quantitative formulation of the  above picture which will lead to an evaluation of the analogs of thermodynamic  properties of a classicalon like entropy and temperature and also a computation of the number of its decay products. We will then use these results to make predictions for signals at the LHC.  As we will see, like black holes, classicalons  decay to give high multiplicity final states. Unlike black holes, however, the classicalons do not couple universally to all SM particles. In particular,  there is no direct coupling to light quarks so that classicalons have a much lower production cross-section than black holes of the same energy. For the same reason, classicalon production, unlike black hole production, is not the dominant scattering process at energies above the classicalization scale with other SM scattering processes having  a higher cross section. In Section 2, we carry out the statistical mechanical analysis of classicalon decays and use the results we find to make LHC predictions in Section 3. Finally we make concluding remarks in Section 4.

\section{Classicalon statistical mechanics}
We will now describe a more precise formulation of the intuitive picture in Fig.~\ref{brili} and obtain quantitative results.  In theories that exhibit classicalization, in addition to the free lagrangian there are non-linear self-sourcing terms which are important only if the energy $\sqrt{\hat{s}}$ gets localized in a radius $r_*$ given by Eq.(\ref{main}). This leads to the formation of a classical configuration of mass $M=\sqrt{\hat{s}}$ which decays into many particles. 

We will consider a massless classicalizer field $\phi$ and discuss later how our results can be generalized to the massive case. We will assume that the only requirements for forming a classicalon  are
\begin{itemize}
\item{conservation of energy and momentum, that is,
\bea
|\vec{k_1}|+|\vec{k_2}|....+|\vec{k_N}|=M \label{coe}\\
\vec{k_1}+\vec{k_2}....+\vec{k_N}=0
\eea
where $k_i$ are the four-momenta of the incoming particles, }
\item{localization of the energy of the incoming particles  inside the classical radius $r_*$.}
\end{itemize}
 As we will see later,  the conservation of the 3-momentum does not  lead to any constraint as it is automatically satisfied for $N\gg 1$. As the time reverse of every classicalon formation process is a classicalon decay process, this implies that every possible way of choosing a final state respecting the above conditions  gives us an allowed classicalon decay. We will think of the set of four momenta of the incoming/outgoing particles in a particular formation/decay process of a classicalon of a given mass, $M$,  as a \textit{microstate}.  The combinatoric exercise of counting the number of ways of choosing these four vectors such that the energy adds up to the classicalon mass would be very similar to the statistical mechanical analysis of ideal Bose gasses or blackbody radiation. As we will see, however, unlike the case of an ideal gas or blackbody radiation,   the particles here are not  represented by waves confined to a box.  The wave-packets must have a size and shape such that the second condition is satisfied and this leads to a density of states function different from the blackbody radiation case.  The statistical mechanics of classicalons  will thus be  very different from blackbody radiation resulting in  different  thermodynamic relations. We will now see what the condition for localization of the energy inside the  radius $r_*$ tells us about the geometry of the incoming wave-packets.

\subsection{Geometry of  wave-packets}
\label{geomm}
 We will see in this section that in order to localize most of their energy inside the classicalon radius, $r_*$, the incoming  wave-packets in a classicalon formation process (and thus, by time reversal symmetry, the outgoing wave-packets in a classicalon decay process) can have a longitudinal width at most of the order of $r_*$, but  are allowed to  have a much bigger  transverse length, $\sqrt{N} r_*$, where $N$ is the number of incoming particles.  We will not have to take into account the effect of the  classicalizing interaction as we will assume that  if the wave-packets  are able to localize their energy inside the radius $r_*$,  in the absence of a classicalizing interaction, they would form a classicalon in the presence of one.

\begin{figure}[t]
\centering
\hspace{-0.2 in}
\begin{tabular}{cc}
\includegraphics[width=0.4\columnwidth]{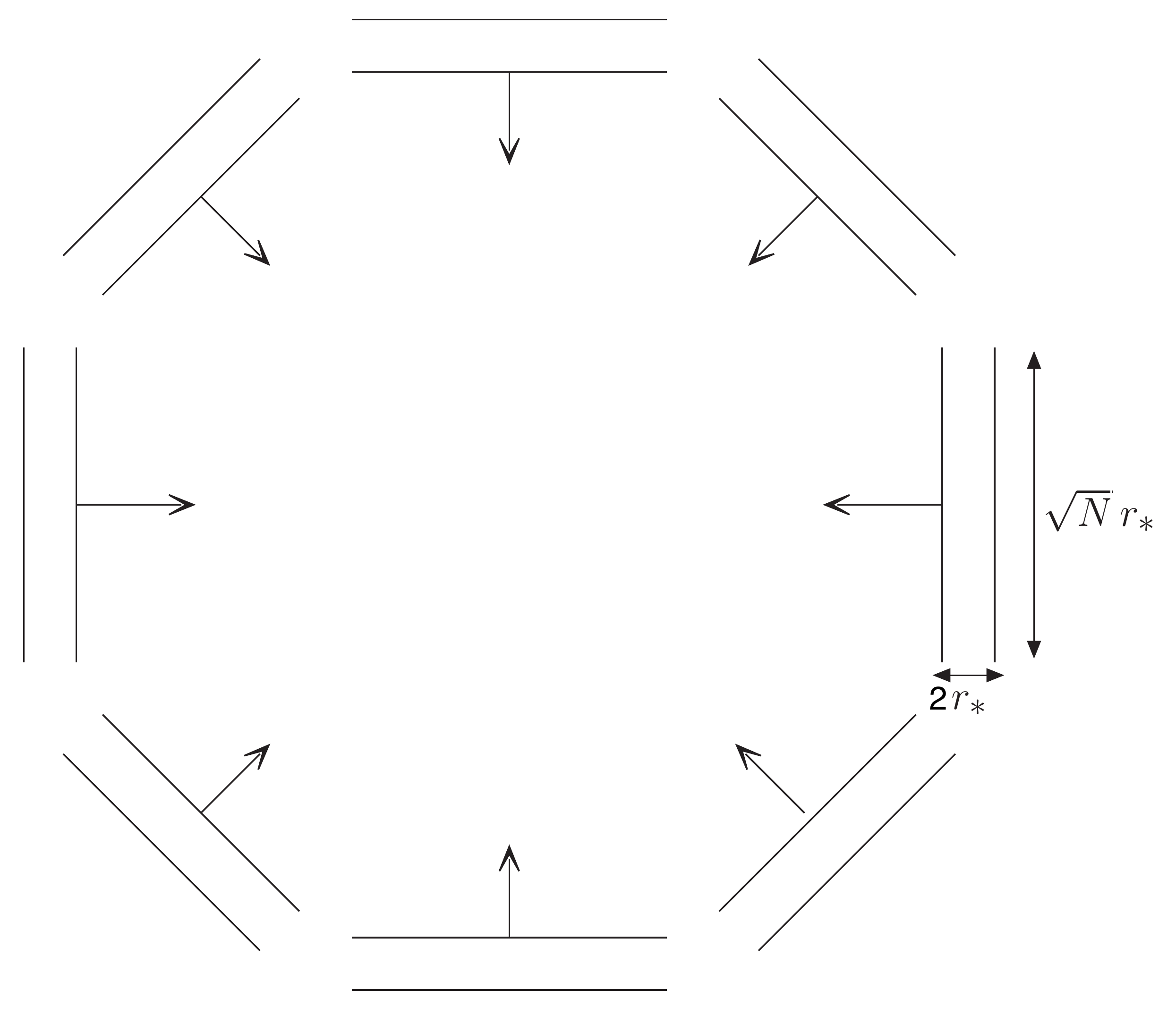}&
\includegraphics[width=0.4\columnwidth]{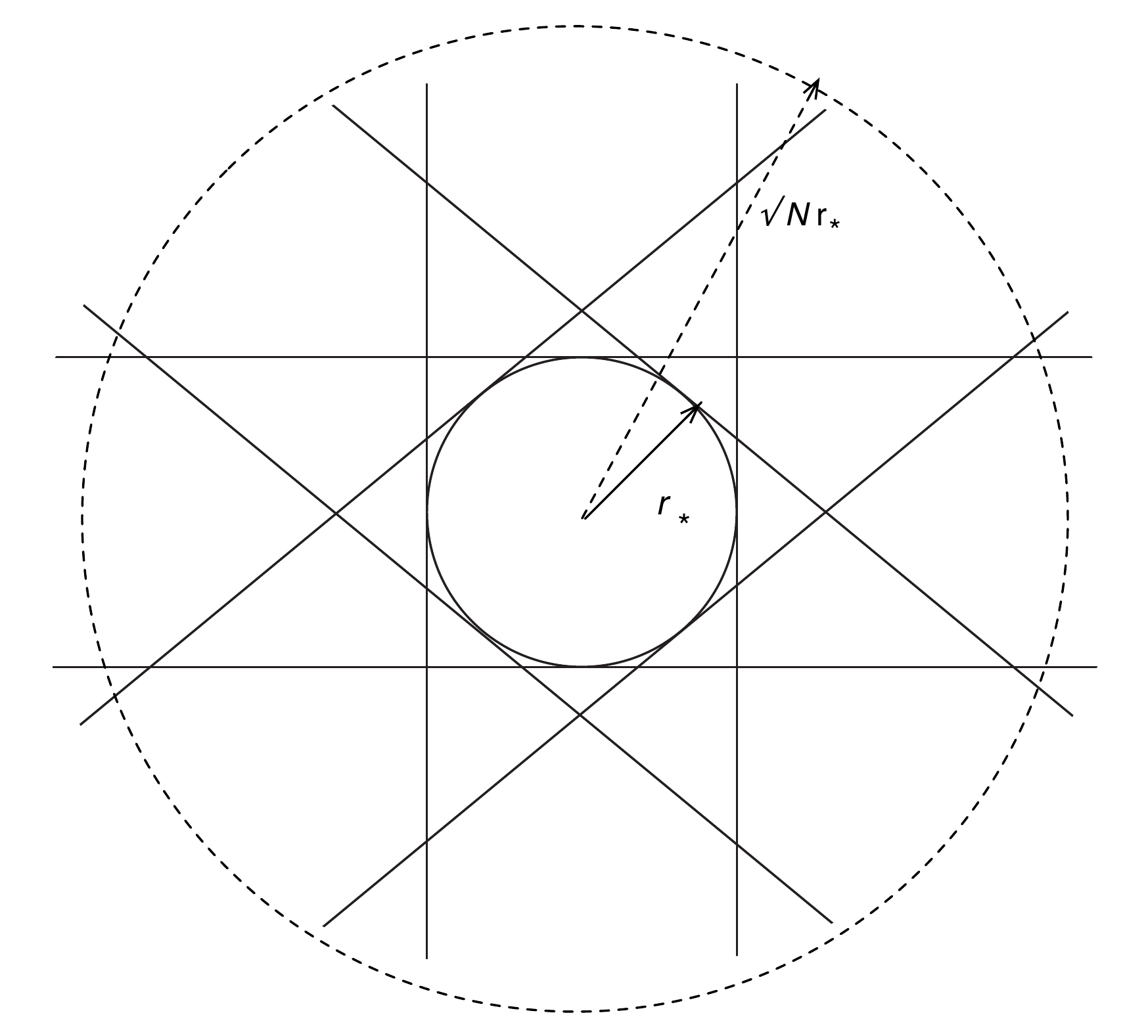}
\end{tabular}

\caption{Classicalon formation from many incoming wave-packets which superpose to give an incoming  spherically symmetric disturbance.   We show the situation at $t\leq0$ (left) and at $t=0$ (right) when the wave-packets reach the origin. We show that at the moment $t=0$ when all the wave-packets reach the origin, a field exists outside the classicalon radius because the wave-packets have transverse length bigger than $r_*$. The field outside, however, drops off as $\phi \sim 1/r$  so that most of the energy is still localized inside $r_*$.  As we discuss in the text, these wave-packets stop overlapping at a distance $\sqrt{N} r_*$  from the origin (shown by the dashed circle here) so we truncate our wave-packets to a length equal to  $\sqrt{N} r_*$ in the transverse directions.    }
\label{tlz}
\end{figure}

We consider the formation of a classicalon from $N$ incoming particles where $1 \ll N \leq N_{max}$ as shown in Fig.~\ref{tlz} (left), propagating freely such that they all reach the origin at the same time, $t=0$. As $N \gg 1$ we can think of these wave-packets to be distributed approximately isotropically in all directions, giving rise to a spherically symmetric incoming disturbance (for $t<0$) when they are superposed with each other. In Ref.~\cite{Dvali:2010ns}  it has been discussed how classicalons can be formed from the collapse of a spherical wave-packet of  finite width. The spherical wave-packet collapses according to the free wave equation when its radius $r>r_*$. As the wave-packet collapses  to a radius smaller than  $r_*$, the non-linear classicalizing term in the lagrangian becomes important and it does not allow the energy to be localized at distances shorter than $r_*$. This leads to the formation of a classical configuration of radius $r_*$ even if the original width of the wave-packet is much smaller.  Clearly the spherical wave-packet cannot have width bigger than the classicalon diameter $2 r_*$ otherwise its energy cannot be localized within the radius $r_*$ and a classicalon would not be formed.

In our picture, such a spherical  disturbance corresponds to a superposition of many incoming `plane' wave-packets of  longitudinal  width $2 r_*$ as shown in Fig.~\ref{tlz} (left). Hence we will take for each wave-packet the boundary conditions for modes confined in a one dimensional box of size $2r_*$.  For an incoming wave-packet with a definite squared energy, $\omega^2=k^2$, we, therefore, take the following functional form in  the longitudinal coordinate $l$,
\beq
\phi(l)=\sin k(l+r_*)
\label{loung}
\eeq
with the  $k$ quantized as,
 \beq
k =n \pi/2 r_*.
\label{quant}
\eeq 
Here $l$ is the longitudinal displacement from the center of the wave-packet and we are not writing the time dependance. Note that the above function satisfies $\phi(l=-r_*)=\phi(l=r_*)=0$.  As any function with compact support  in the width of the wave-packet can be decomposed as a superposition of the above modes, this means that we are considering all possible wave-packet profiles which go to zero outside the width of the  wave-packet. In particular we are considering wave-packets with  widths smaller than $2r_*$.

What about the transverse length of the wave-packets? In the transverse direction the wave-packets can actually have a length much bigger than $r_*$. This leads to the existence of a field outside the classicalon radius $r_*$ when the wave-packets superpose at the origin at $t=0$, as is clear from Fig.~\ref{tlz}(right), but, as we show in Appendix A,  the field outside the classicalon radius drops off as $\phi \sim 1/r$ so that most of the energy is still inside the classicalon radius $r_*$. The $1/r$ behavior is  expected because we are superposing solutions of the free wave equation which becomes Poisson's equation in the static limit. For $t>0$, the classicalon decays and it is clear that the field at any point outside $r_*$ remains unchanged from its $t=0$ value  until the information of the classicalon decay reaches it. Hence if there is a $\phi \sim 1/r$ tail at $t=0$,  we expect such a tail to remain at points outside, till the information of the classicalon decay reaches them. Similarly there exists a $\phi \sim1/r$ tail outside the incoming wave-packets for $t<0$. The appearance of the  $\phi \sim1/r$ tail outside the classicalon is interesting because such a tail in fact exists in the static solutions discussed in Ref.~\cite{Dvali:2010jz}. It is necessary because it leads to the flux of the gradient $\nabla \phi$ that must exist because of the source. We will show in Appendix A that $\phi \sim Q/r$, where $Q=\sqrt{N}$ matches with the `charge' of the classicalizing source at the parametric level.

So far we have been assuming that the wave-packets are infinitely large in the transverse direction. This would, however, create a problem unless we have a superposition of an infinite number of wave-packets. This is because, as it is clear from Fig.~\ref{tlz}(right), if there are a finite number of wave-packets, at large distances the wave-packets will not overlap anymore and thus we would not get the superposition leading to the $1/r$ fall off of the field. For a finite number of wave-packets with infinite transverse dimensions  most of the energy of the wave-packets would be  localized at large distances where there is no overlap between the different wave-packets. Thus our wave-packets must have large but finite transverse dimensions. We show in Appendix A that the distance at which the wave-packets stop overlapping is given by,
\beq
L=\sqrt{N}r_*.
\label{overlap}
\eeq
Thus we see that the incoming/outgoing particles in a classicalon formation/decay process, can be represented by wave-packets of size $2 r_*$ in the longitudinal direction and size $\sqrt{N} r_*$ in the transverse direction.  

Before going into our quantitative derivations, we will describe what happens qualitatively. At times $t<0$ and  distances from the origin much larger than   $r_*$, the wave-packets travel freely and the number of quanta is conserved.  As the wave-packets approach distances closer than $r_*$, the non-linear classicalizing term becomes important, the number of particles is no longer conserved and can increase or decrease from the initial number. As we said earlier, we will think of the set of four-momenta in a particular formation/decay process of a classicalon of a given radius as a microstate.  Whereas the initial number of particles and their momenta can be arranged to be anything by us, we would expect the classicalon to decay to a  number of particles and with an  energy distribution for the decay particles that corresponds to the maximum number of microstates. We want to find this distribution function that corresponds to maximum number of microstates. The first ingredient we need is the density of states function.

\subsection{Density of states function}
\label{dos}
We want to find out the density of states for the wave-packets we described, that is the number of  wave-packets of the kind described above that have energy in the range  $\omega$ to $\omega + d \omega$. We will obtain such wave-packets by superposing free wave modes confined in a box of volume $V=L^3$ where $L$ is given by Eq.(\ref{overlap}). To get a wave-packet with momentum $\vec{k}$ and width $2 r_*$ we would have to superpose many waves with momentum in the same direction as  $\vec{k}$ and magnitude around $|\vec{k}|$.\footnote{Note that this way of constructing our wave-packets ensures that any two wave-packets traveling in different directions are linearly independent. The    functional form in Eq.(\ref{loung}) ensures that two wave-packets in the same direction, but with different $\omega = k$, are also linearly independent.}   For waves confined in a box all values of $(k_x,k_y,k_z)$ are not allowed, instead only a lattice of points  in $k$-space is allowed. Another way of saying this is that in a shell in $k$-space between the radii $\omega$ and $\omega + d \omega$ all possible directions are not allowed. We want to find the number of states that lie within this shell. For the modes confined in the box we know that the density of states is given by,
\beq
g({\vec{k}}) d^3k=\frac{  V }{8 \pi^3} d^3 k.
\eeq
Going to spherical coordinates, $d^3 k \to 4 \pi k^2 dk= 4 \pi \omega^2 d \omega$, this gives,
\beq
g(\omega) d\omega=\frac{  V \omega^2}{2 \pi^2} d\omega=\frac{1}{8\pi^3} L d\omega \times  L^2 (4 \pi\omega^2)
\label{omk1}
\eeq
Up to factors of $\pi$ the first term here is the number of box modes  in a particular direction having energy in the range $\omega$ to $\omega + d \omega$ and the second term is the number of allowed directions. As we are considering wave-packets of width $ 2r_*$ and not $L$ in the longitudinal direction, the number of wave-packets in a particular direction with energy in the range $\omega$ to $\omega + d \omega$ will be smaller by a factor $2 r_*/L$ so that we get,
\beq
g(\omega) d\omega=\frac{1}{8\pi^3} (2 r_*) d\omega \times  L^2 (4 \pi\omega^2)= \frac{N r_*^3 \omega^2}{\pi^2} d\omega
\label{omk2}
\eeq
where we have substituted $L$ from Eq.(\ref{overlap}). One must also keep in mind that there are no wave-packets with energy  less than $ \pi/ 2r_*$ (see Eq.(\ref{quant})). It is useful to write the density of states function also in cartesian coordinates,
\beq
g({\vec{k}})~ d^3 k= \frac{N r_*^3 }{4 \pi^3}~ d^3 k.
\label{cdos}
\eeq
Note that the existence of the extra factor of $N$ in Eqs.(\ref{omk2}) and~(\ref{cdos}), as compared to the case of a particle confined in a box, is crucial and leads to thermodynamic relations for a classicalon different from ideal Bose gasses or blackbody radiation.
\subsection{Number of  $N$ particle decays for $1\ll N \ll N_{max}$ }
We want to count the number of ways a classicalon of mass $M$ can decay to $N$ particles which is the same as the number of ways of forming a classicalon from $N$ particles. We want to show that the number of ways is higher for larger $N$, thus proving that a classicalon prefers to decay to many particles.  We will now evaluate $\Gamma(M,N)$, the number of ways in which $N$  incoming particles, where  $1\ll N \ll N_{max}$, can form a classicalon of a given mass, $M$. Note that for our derivation here we will assume that  in each energy state there is at most one particle which is a very good approximation  for $N\ll N_{max}$. The total number of ways of forming a classicalon would be,
\beq
\Omega(M) =\sum_{N=2}^{N_{max}}\Gamma(M,N).
\label{boltz}
\eeq
We try to find all possible set of four vectors of the $N$ outgoing wave-packets with the only constraint that energy and momentum are conserved,
\bea
|\vec{k_1}|+|\vec{k_2}|....+|\vec{k_N}|=M \label{coe}\\
\vec{k_1}+\vec{k_2}....+\vec{k_N}=0
\eea
For large $N$, the momentum conservation constraint is not  important. This is because the sum $\vec{k_1}+\vec{k_2}....+\vec{k_{N-1}}$ is completely unconstrained as we can always fix $\vec{k_N}$ to ensure that the sum $\vec{k_1}+\vec{k_2}....+\vec{k_{N}}=0$. For $N \gg 1$, $|\vec{k_N}|$, which is the energy of a single particle is  negligible compared to $M$, so that the two conditions above can be reduced to a single energy conservation condition $|\vec{k_1}|+|\vec{k_2}|....+|\vec{k_{N-1}}|=M$ on the $N-1$ particles. For large $N$, however we can always replace $N-1$ by $N$. Using Eq.(\ref{cdos}) we thus get the following phase space integral with only the energy conservation constraint,
\beq
\Gamma(M,N)=\frac{((N/4) (r_*/\pi)^3)^N}{N!} \int d^3 k_1 d^3 k_2...d^3 k_N \delta(|\vec{k_1}|+|\vec{k_2}|..+|\vec{k_N}|=M).
\eeq
The $N!$ in the denominator appears because the particles are indistinguishable and all possible permutations result in the same state.   The integral above is a well-known integral in statistical mechanics that appears in the evaluation of entropy of an ideal ultra-relativistic gas. For $N \gg 1$, the result is (see for instance Ref.~\cite{greiner}),
\beq
\Gamma(M,N)=\frac{2^{N} (\sqrt{3})^{3N}N^N(r_* M/ \pi)^{3N}}{N!(3N)!}=\frac{2^{N} (\sqrt{3})^{3N}N^N N_{max}^{3N}}{\pi^{3N} N!(3N)!}.
\eeq
 It is easy to check that  $\Gamma(M,N)$ is an increasing function of $N$,
which shows that a classicalon would prefer to decay to many particles and not a few.

\subsection{Classicalons as Bose-Einstein systems}
In this subsection we will try to find the most probable energy distribution of the particles a classicalon decays to.  In other words we will try to find the distribution with the maximum number of microstates, $\tilde{\Omega}(M)$. As is usually assumed in statistical mechanics we will assume that the total number of  ways of forming the classicalon, $\Omega(M)$ in Eq.(\ref{boltz}), is approximately equal to the total number of ways of forming the most probable distribution, that is,
\beq
\Omega(M) \approx \tilde{\Omega}(M).
\eeq
Parts of the discussion here will be very similar to the standard derivation of the Bose-Einstein distribution,  although the density of states function here is different.  

We want to find the most probable value of  $N_\omega$, the number of particles in the energy state with energy  $\omega$.  In the continuum limit, $N_\omega$ becomes $N(\omega)$, the distribution function.  As explained in the previous subsection,  the only constraint is  the energy conservation constraint  in Eq.(\ref{coe}) which we rewrite  as,
\beq
\sum_\omega N _\omega g_\omega \omega~ d \omega=M.
\label{cont2}
\eeq
where $g_\omega$ is the degeracy of the energy state with energy $\omega$. In the continuum limit, $g_\omega$ becomes $g(\omega)$, the density of states function derived in Section~\ref{dos}. We first need to find  $\Omega(M)$,  the number of  ways of choosing the four momenta of the decaying particles  while satisfying the constraint in Eq.(\ref{cont2}). As we review in Appendix B, this is given by the well known expression,
\beq
\Omega(M)=\Pi_\omega \frac{(N_\omega
+g_\omega)!}{N_\omega! g_\omega!}
\eeq
We can define the entropy of the system  as,
\beq
S=\log (\Omega(M)).
\eeq
We want to maximize $S$ respecting the constraints in Eq.(\ref{cont2}). As shown in Appendix B, using the method of Langrange multipliers, this leads to the Bose-Einstein distribution,
\beq
N_\omega= \frac{g_\omega}{e^{\beta \omega}-1}.
\eeq
Here $\beta$ is the Lagrange multiplier related to the constraint in Eq.(\ref{cont2}) and effectively plays the role of  inverse temperature, $T^{-1}$.  To obtain $\beta$ and the number of particles in the most probable distribution, $N_*$, we now go to the continuum limit replacing the summations above by integrals and  solve the equations,
\bea
N_*= \int_{\omega=\pi/ 2r_*} g(\omega)d \omega=\frac{N_* r_*^3 }{ \pi^2}\int_{\omega=\pi/ 2r_*}\frac{\omega^2 d\omega}{e^{\beta \omega }-1}
\label{cons1}\\
M= \int_{\omega=\pi/ 2r_*} \omega g(\omega)d \omega=\frac{ N_* r_*^3 }{\pi^2}\int_{\omega=\pi/2r_*}\frac{\omega^3 d\omega}{e^{\beta \omega }-1}.
\label{cons2}
\eea
Note that the lower limit in the integral is not zero but the minimum allowed frequency for our wave-packets $\pi/2r_*$  (see Eq.(\ref{quant})). To obtain $\beta$ make the substitutions $\beta \omega = x$  in Eq.(\ref{cons1}) to obtain,
\beq
\int_{x=\beta \pi/2r_*}\frac{x^2 dx}{e^{x}-1}=(1/\pi)(\beta \pi/r_*)^{3}.
\label{125}
\eeq
Both the LHS and RHS of the above equation depend on $\beta \pi/r_*$. While the RHS obviously increases with $\beta \pi/r_*$ the integral in the LHS decreases as the lower limit is raised so  it decreases with $\beta \pi/r_*$ and we find a unique solution at $\beta \pi/r_* \approx1.9$. The precise numerical coefficients should not be taken seriously and only the parametric relationships are important. We get\footnote{As the radius of the classicalon increases with mass (see Eq.~(\ref{main})), this relationship implies that the classicalon temperature decreases with its energy  so that it has a negative specific heat.},
\beq
\beta\sim r_*\Rightarrow T\sim \frac{1}{r_*}.
\label{betr}
\eeq 
Now to find $N_*$ we use Eq.(\ref{cons2}), again substituting $\beta \omega = x$, to get,
\beq
\frac{ N_* r_*^3 }{ \pi^2}{\beta^{-4}}\int_{x=\beta \pi/2r_*}\frac{x^3 dx}{e^{x }-1}=M.
\label{126}
\eeq
Now substituting the solution of Eq.(\ref{125}), $\beta \sim r_*$, we get,
\beq
N_*\sim {M r_*}\sim N_{max}.
\eeq
The above expression shows that the typical energy of a quanta is $M/N_* \sim 1/r_*$, so that the typical wavelength is $r_*$. This is what we  expect from the dynamics of classicalization as $r_*$ is the length scale at which classicalization takes place~\cite{Dvali:2011th}. Now we can also evaluate the entropy,
\bea
S&=&\int \beta dM \sim \int r_* dM \sim \int\frac{M^{~\alpha}}{M_*^{1+\alpha}} dM \sim \left( \frac{M}{M_*}\right)^{1+\alpha}\sim M r_* \nonumber\\
&\Rightarrow& S \sim N_*\sim N_{max}
\eea
where we have substituted $r_*$ using Eq.(\ref{main}) taking $\sqrt{\hat{s}}=M$. Thus we have found  that  the classicalon decays to the maximum number of particles it can, $N_{max}$, with a blackbody spectrum having $T\sim 1/r_*$. We see that the total number of decays is $\Omega(M)=e^S$, so that probability of decays to a few particles, which is a small number compared to $e^S$, would be exponentially suppressed,
\beq
P(Classicalon \to few)\sim \frac{1}{\Omega(M)}\sim e^{-S} \sim e^{-N_{*}}.
\eeq
in accordance with Ref~\cite{Dvali:2011th} .

We now consider the special case of a black hole for which $\alpha=1$, $M_*=M_{pl}$ and $r_*$ is the Schwarzschild radius. As a black hole does not decay classically, the above analysis for the distribution of the decay products cannot be applied to a black hole. As argued in Ref.~\cite{Dvali:2011th}, however, classicalization is the first step to the formation of a black hole and this takes place  before the horizon emerges. Thus our calculation of the entropy which is basically a counting of the number of ways in which a classicalon can be formed should give us the correct black hole entropy. Indeed, we find for $\alpha=1$,
\beq
S \sim N_* \sim M r_* \sim M_{pl}^2 r_*^2
\eeq
in agreement with the Bekenstein-Hawking formula.

We want to emphasize that obtaining the parametric relationships above is far from assured based only on dimensional grounds. For instance if we had taken wave-packets of size $2r_*$ in both the longitudinal and transverse directions we would have found the usual  density of states for an ideal gas, $g(\omega) d\omega\sim r_*^3 \omega^2 d\omega$ without the factor $N$. This would lead to the usual relationships $M\sim  r_*^3 T^4$ and $S\sim r_*^3 T^3$ for blackbody radiation. The fact that the wave-packets are of size $\sqrt{N }r_*$ in the transverse direction is thus crucial in obtaining the  final result we have derived. As we discussed earlier in Section~\ref{geomm} (and show in detail in Appendix A) a transverse length much greater than $r_*$ is in   fact necessary for generating the $\phi \sim 1/r$ tail of the field outside the classicalon. 

In the above analysis we have considered  all frequencies higher than $\omega=\pi/2r_*$, including frequencies higher than  the  cut-off $M_*$, although the  higher frequencies are exponentially suppressed by the Bose-Einstein distribution function.  We have not imposed an energy cut-off $M_*$ disallowing higher frequencies or equivalently wavelengths smaller than the  length cut-off $L_*=1/M_*$. Such a cut-off would be analogous to the Debye frequency of crystals. Note that such an energy cut-off if imposed will not make any difference  to our final results in the classical (and thermodynamic) limit $\sqrt{s}/M_*\to \infty$, where $\sqrt{s}=M$. To see this note that an upper cut-off  $\omega=M_*$ in the integrals in Eq.(\ref{cons1}) and (\ref{cons2}) corresponds to a cut-off $\beta M_*$ in the integrals in Eq.(\ref{125}) and (\ref{126}). Now,
\beq 
\beta M_* =r_* M_* =\left(\frac{\sqrt{s}}{M_*}\right)^{\alpha} \to \infty
\eeq
as $\sqrt{s}/M_*\to \infty$ where we have used  Eq.~(\ref{main}) and (\ref{betr}).  Thus, in the classical limit, putting such an ultraviolet  cut-off is equivalent to putting no cut-off at all and hence it will not change our results.

We want to mention some modifications  that we will make in our  expressions before using them for experimental predictions The first issue is regarding the lower limit $\omega=\pi/2r_*$ in the integrals. It is not true that energies smaller than $\omega=\pi/2r_*$, or larger wavelengths, $\lambda \gg r_*$, are not present. This is because the distribution function we have  derived is for wave-packets of size of the order of $ r_*$ in the longitudinal direction. A detector, however, would detect plane waves much larger in size and the  wave-packets of size $r_*$ are themselves composed of  plane waves of much  have larger wavelengths. Note that this does not have any effect on the distribution function for  higher energies (smaller wavelengths). We will not attempt to find the correct distribution at lower energies (longer wavelengths) as our assumption that the distribution function suddenly drops to zero for wave-packets with width larger than $r_*$  is a simplifying approximation and is not accurate. It is reasonable to expect that a more precise analysis would yield the Planck distribution over the whole energy range. Even in this case the lower frequencies would be  suppressed  due to the phase space factor $\omega^2 d\omega$. Therefore from here onwards we will get rid off the lower limit $\omega =\pi/2 r_*$  in the integrals and take the  lower limit to  be the lowest kinematically allowed value, $\omega=m$, $m$ being the mass of the $\phi$-quanta. We need to make a second modification because we have been assuming so far,  a classicalizer field that is massless,  whereas it is massive in the models we are going to consider. While deriving our density of states function we made in Eqs.(\ref{omk1}) and~(\ref{omk2}) the substitution $k^2 dk= \omega^2 d \omega$ which assumes that the $\phi$-quanta are massless. Using $k^2= (\omega^2-m^2)$ instead, $m$ being the mass,  we  get the correct density of states expression in the massive case,
\beq
g(\omega) d\omega \sim N r_*^3   k ^2 dk\sim N r_*^3   \omega \sqrt{\omega^2-m ^2} ~d\omega.
\label{mass}
\eeq
Finally, in order to make experimental predictions, we will fix the unknown numerical coefficients in the parametric form of the density of states function above by using the black hole example where the exact expressions are well known. We will describe this in more detail in the next section.
\label{two}

\section{Classicalons at the LHC}

Now that we know how to compute the number of decay products in a classicalon decay,  we are ready to perform a collider study  of  classicalization in the phenomenological models introduced in the Section~\ref{one}. Along with the decay multiplicity computation, the other important fact that we will use for our study is that  classicalon production has a geometric cross-section $\pi r_*^2$.   The two models we are going to consider are the classicalization of longitudinal $W$s and $Z$s and the classicalization of Higgs bosons. The LHC signal would be multi-$W/Z$ production in the first case and multi-Higgs production in the second case. As in the case of black hole production in TeV-scale quantum gravity models~Ê\cite{Antoniadis:1998ig, Dvali:2001gx, Giddings:2001bu, Dimopoulos:2001hw}, this would finally lead to production of leptons and many jets. Unlike black holes, though, classicalon production would not be a universal phenomenon in hard scattering processes at energies above the cut-off scale.\footnote{In gravitational high energy scattering above the Planck scale, black hole formation is expected for impact parameters smaller than the Schwarzschild radius or equivalently for large scattering angles. For  impact parameters  much larger than the Schwarzschild radius and transplanckian energies elastic 2$\to$2 scattering should take place which is well described by the eikonal approximation ($t/s \ll 1$)~\cite{Giudice:2001ce, Stirling:2011mf}.}  This is because the light quarks and gluons would not have a strong coupling to the classicalon in both the cases we will consider. This is the main difference of classicalization signals from black hole signals. Thus even at energies higher than the classicalization scale, normal SM $2 \to 2$ hard scattering processes would continue in other channels with  cross-sections larger than classicalon production. Another result of the absence of  any direct coupling between classicalons and light quarks or gluons is that  classicalon production  would have a much smaller cross-section compared to black hole production at the same scale, so that classicalons would be harder to discover/exclude at colliders.

Before going into the details,  there is a caveat that must be emphasized. The phenomenon of classicalization is well understood  only for energies much higher than the classicalization scale. In particular the quantitative expressions that we will use, for instance, the expressions for the radius, cross-section and decay multiplicity, strictly hold only in the limit of large number of quanta, i.e. for $N_* \gg 1$ or equivalently for energies much higher than  the cut-off, $\sqrt{\hat{s}}\gg M_*$. This is the classical limit as well as  the thermodynamic limit where our statistical assumptions are true. As the energies accessible at the LHC are not so high, we will be forced to consider  processes where $N_* \sim 6$. Many would  consider these energies to be still part of the  `quantum regime' around the classicalization scale. The same problem exists in collider analyses of black hole formation and decays in TeV-scale quantum gravity scenarios~\cite{Meade:2007sz}. Black holes can be reliably tackled by theory only at energies much higher than the Planck scale. In the  regime around the classicalization (Planck) scale, it is more appropriate to think of classicalons (black holes)  as a tower of quantum resonances than as classical objects~\cite{Dvali:2011nh}. There is, however, no theoretical model for this quantum regime that can be used to make reliable experimental predictions. Thus in the absence of a better alternative the only choice we have is to use the expressions for the classical regime, as has been done in studies of black holes so far. We will, however, incorporate  in our analysis the fact that classicalon masses are  quantized.

\subsection{Classicalization of longitudinal $W$s and $Z$s}

As is well known, in the absence of the Higgs boson the scattering of  the longitudinal components of $W$ and $Z$ bosons violates tree-level unitarity at energies of the order of a TeV~\cite{Lee:1977eg}. In Ref.~\cite{Dvali:2010jz} it was proposed that classicalization can unitarize these amplitudes. In this proposal the longitudinal (goldstone) modes of the vector bosons classicalize and form a configuration of $W$s and $Z$s that finally decay into many $W$s and $Z$s. 

 For our anlysis we will  take the classicalizing interaction proposed in  Ref.~\cite{Dvali:2010jz},
\beq
\frac{c}{2} ({\rm Tr}\left(D^\mu U D_\mu U^\dagger\right))^2
\eeq
where $U$ is the SU(2) matrix $U= \exp(i \pi_a \tau_a/v)$ containing the goldstones $\pi_a$. Here $v=246$ GeV is the Higgs vacuum expectation value (VEV) and $\tau_a$ are the Pauli matrices. The covariant derivative  above  is defined as follows,
\beq
D_\mu U =\partial_\mu U  +ig \frac{\tau_a}{2} W^a U - i g'U B_Y \frac{\tau_3}{2}.
\label{cov}
\eeq
When expanded the operator in Eq.(\ref{cov}) gives the following classicalizing interaction,
\beq
\frac{c}{v^4}(\partial_\mu \pi_a \partial^\mu \pi^a)^2.
\eeq
For this particular operator the classicalon radius is given by~\cite{Dvali:2010jz, Dvali:2010ns},
\beq
r_* \sim c^{1/3} \frac{M^{1/3}}{v^{4/3}}\sim  \frac{M^{1/3}}{M_*^{4/3}},
\label{unnorm}
\eeq
where, $M_* = v/c^{1/4}$, is the classicalizing scale. Note that the above relationship is valid only until $r_*$ reaches the Compton wavelength of the $Z$-boson, $1/m_Z$. Beyond this point the radius would freeze at the value $1/m_Z$~\cite{Dvali:2010jz}. The experimental constraints on the coupling $c$ come from electroweak precision measurements. Only  the $T$ parameter gets a contribution from this operator and the other electroweak parameters (the $S$ parameter and the six electroweak parameters $U$-$Z$  as defined in  Ref.~\cite{Barbieri:2004qk}
 for instance) get no contribution. The contribution to the $T$  parameter is given by~\cite{Dutta:2007st},
\beq
\Delta T= \frac{-(c/2)}{4 \pi^2 \alpha_{em}}\left( \frac{3 g^2 g'^2}{2}+\frac{3 g'^4}{4}\right)\log \frac{M_*}{M_Z}.
\label{delt}
\eeq
 As $c=(v/M_*)^4$, we see that the $\Delta T$ contribution is small for $M_* \gtrsim 500$ GeV.  For $M_*= 246$ GeV the contribution is appreciable. From Eq.(\ref{delt}) we find that for $c=\pm1$(and hence $M_* = v$) we get $\Delta T= \mp 0.1$. As we are considering a higgsless theory a negative $c$ is preferred. For higher values of   $M_*$ the contribution to $\Delta T$ would be much smaller.   There would, however, be additional contributions to electroweak precision observables from the quantum resonances that  exist in such a theory around the classicalizing scale, $M_*$. These  contributions are unfortunately not calculable  without a knowledge of the precise dynamics at the classicalizing scale. All we can do is make the general statement that a higher classicalization scale will  mean smaller contributions to electroweak precision observables from these resonances.

 We will absorb the unknown numerical coefficient in Eq.(\ref{unnorm}) in a redefinition of the coupling $c$ to obtain,
\beq
r_* =c^{1/3} \frac{M^{1/3}}{v^{4/3}}=  \frac{M^{1/3}}{M_*^{4/3}}.
\eeq
Note that the classicalization  scale, $M_*=v/c^{1/4}$, cannot be much higher than the TeV scale as $WW$-scattering needs to be unitarized before these energies are reached. We will make computations for the three choices of the classicalization scale,  $M_*=246$ GeV, $M_*=600$~\textrm{GeV} and $M_*=1$ TeV.

\subsubsection{Multiplicity of gauge bosons in the final state}

We want to find  the total number of $W/Z$s a classicalon of a given mass, $M$, would finally decay into. We are not allowed to use the massless limit of the expressions we derived (Eqs.(\ref{cons1}) and~(\ref{cons2})) in this case. One way of seeing  that the massless approximation is not valid here is that  the expression for multiplicity in the massless limit would give us multiplicity greater than the kinematic bound $M/m_{W/Z}$.  The reason we need to consider the mass is that, in this case, the kinetic energy $k\sim1/r_*$ does not dominate the energy of the individual quanta as the mass of the quanta is comparable, that is  $m_{W/Z}\sim 1/r_*$. This is in turn because of the small separation between the mass $m_{W/Z}$ and the classicalization scale $M_*=v/c^{1/4}$. Thus we use the the density of states for the massive case given previously in Eq.(\ref{mass}),
\beq
g(\omega) d \omega=\gamma {N_* r_*^3 } \omega \sqrt{\omega^2-m^2} d\omega,
\eeq
where $\gamma$ is an unknown numerical coefficient that we will fix by demanding that we get the exact result in the black hole case.  We will not consider here the effects of the difference in $W$ and $Z$ mass which is small compared to the classicalon mass. To be conservative we will take the mass of all the quanta forming the classicalon, $m=91.2$ GeV the $Z$-mass. We find the number of decay particles by solving for $\beta$ and $N_*$, the Eqs.(\ref{cons1}) and~(\ref{cons2}) but with the modified density of states function in Eq.(\ref{mass}) and a different lower limit,
\bea
\gamma {N_* r_*^3 }\int_m^M \frac{\omega \sqrt{\omega^2-m^2} d\omega}{e^{\beta \omega }-1}=N_* \label{consm1}\\
\gamma{ N_* r_*^3 }\int_m^M \frac{\omega^2  \sqrt{\omega^2-m^2} d\omega}{e^{\beta \omega }-1}=M.
\label{consm2}
\eea
We have explained at the end of Section~\ref{two} why the lower limit in the  integrations above has been changed  from the lower limit in Eqs.(\ref{cons1}) and~(\ref{cons2}). We fix the factor $\gamma$ above by requiring  that for $m=0$ we get from Eq.(\ref{consm1}),  the exact black hole result,
\beq
\beta^{-1}=T=\frac{1}{4 \pi r_*}.
\label{num}
\eeq
This gives\footnote{Note that the analytical expression strictly holds only in the limit that the upper limit  (after the substitution $\beta \omega=x$) of the integration in Eq.(Ê\ref{consm1}) , $\beta M \to \infty$. We, however, find it to be a very good approximation in the examples we consider.},
\beq
\gamma=\frac{(4 \pi)^3}{2 \zeta(3)} \approx 825,
\label{norm}
\eeq
where $\zeta(n)$ is the Riemann zeta function.
\begin{figure}[t]
\centering
\hspace{-0.3 in}
\begin{tabular}{cc}
\includegraphics[width=0.47\columnwidth]{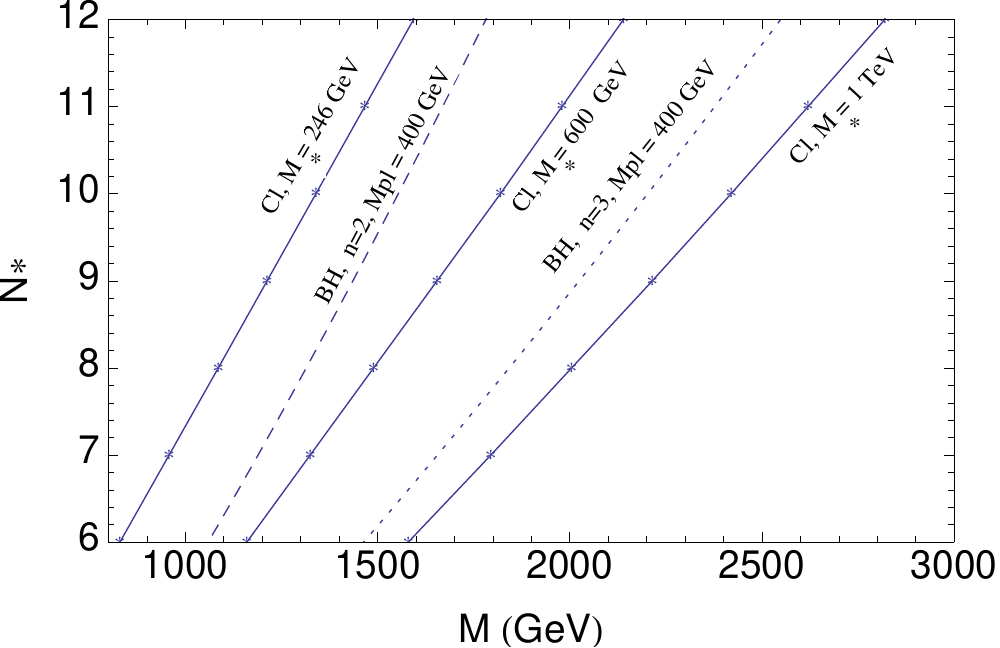} &
\includegraphics[width=0.47\columnwidth]{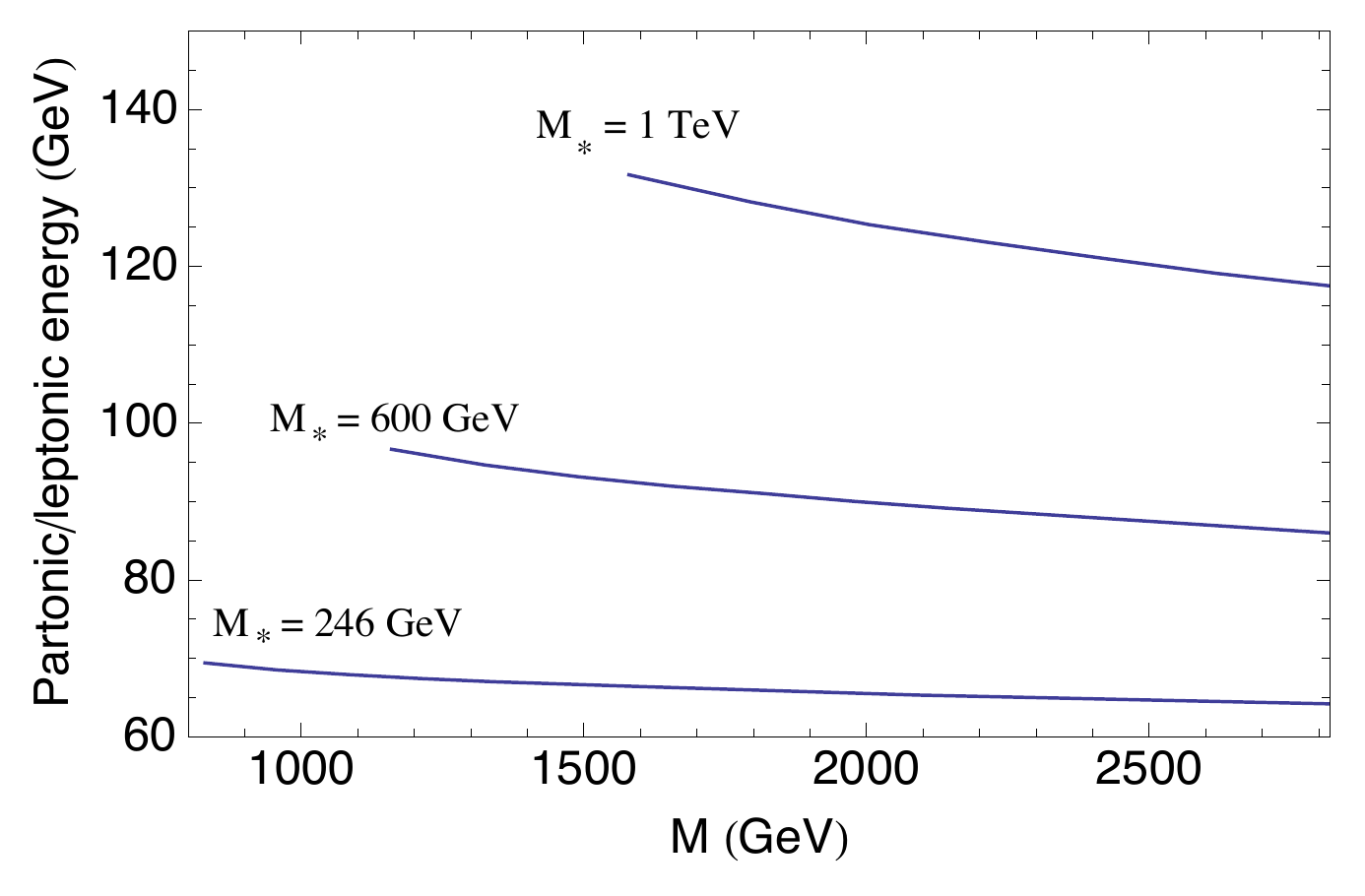} 
\end{tabular}
\caption{In the  model of classicalization of  longitudinal $W$s and $Z$s we show the number of quanta $N_*$ as a function of the classicalon mass $M$ in the  figure on the left. In the figure on the left,  we also show the $N_*$ vs $M$ curves for some black hole examples. Classicalon states exist only at the points where values of $N_*$ are integers. In the figure on the right, we show the typical energy, $M/2N_*$, of a lepton or parton emerging from one of the $W/Z$s produced in the classicalon decay, in the rest frame of the classicalon. }
\label{nfig}
\end{figure}

The results of our evaluation are shown in Table~\ref{bigt} and Fig.~\ref{nfig} (left) for our three choices, $M_*=246$ GeV, $M_*=600$ GeV and $M_*=1$ TeV. We see that instead of the dependence $N_*\sim M r_*\sim M^{~4/3}$ expected in the massless limit, we find an almost  linear dependence $N_*\sim M$ (the dependence is not exactly linear as can be seen from the values in  Table~\ref{bigt}).   For comparison we also show the $N_*$ vs  $M$ dependence for  extra-dimensional black holes in Fig.~\ref{nfig} (left). We have used the expression for $N_*$  in Ref.~\cite{Dimopoulos:2001hw},
\beq
N^{BH}_*=\frac{2 \sqrt{\pi}}{n+1}\left(\frac{M}{M_{pl}}\right)^{\frac{n+2}{n+1}}\left(\frac{8 \Gamma(\frac{n+3}{2})}{n+2}\right)^{\frac{1}{n+1}}.
\label{bh}
\eeq
Here $n$ is the number of extra dimensions and $M_{pl}$ is the fundamental Planck scale in the $4+n$ dimensional space-time. We have taken $n=2, 3$ and $M_{pl}=400$ GeV. Note that the value $M_{pl}=400$ GeV has been chosen close to the classicalization scale only for comparison and such low values of $M_{pl}$ have already been ruled out~\cite{BHpas}. Higher values of $M_{pl}$ will give much lower $N_*$ values. As one can see in  the figure, for the $n=2$ case, the $N_*$ vs $M$ curve is clearly not linear whereas for the $n=3$ case the non-linearity due to the $N_*\sim M^{\frac{n+2}{n+1}}$ dependence is not noticeable.  As $n$ is increased (note that larger $n$ values are preferred because of  astrophysical bounds~\cite{Nakamura:2010zzi}) the curve would become more and more linear and  $N_*$ would   decrease. Note that whereas $N_*$ is the final decay multiplicity in the case of black holes,  in the case of classicalons the multiplicity of final decay products is actually bigger  than (about twice) $N_*$, because $N_*$ is just the number of the primary decay products, the $W$s and $Z$s, which decay further giving rise to  more leptons and jets.   Keeping this in mind one can from see from Fig.~\ref{nfig} (left) that  the multiplicity of final decay products is  larger for these classicalons when compared to black holes of the same mass even for  such small values of $n$ and $M_{pl}$ as $n=2$ and $M_{pl}=400$ GeV. In Fig.~\ref{nfig} (right) we show the typical energy, $M/2N_*$, of a lepton or partonic jet emerging from one of the $W/Z$s produced in the classicalon decay, in the rest frame of the classicalon. An experimental  measurement of the typical energy would tell us about the $N_*$ vs $M$ dependance for the classicalon. We will discuss this measurement in more detail later. 
\begin{figure}[t]
\centering
\hspace{-0.3 in}
\begin{tabular}{cc}
\includegraphics[width=0.47\columnwidth]{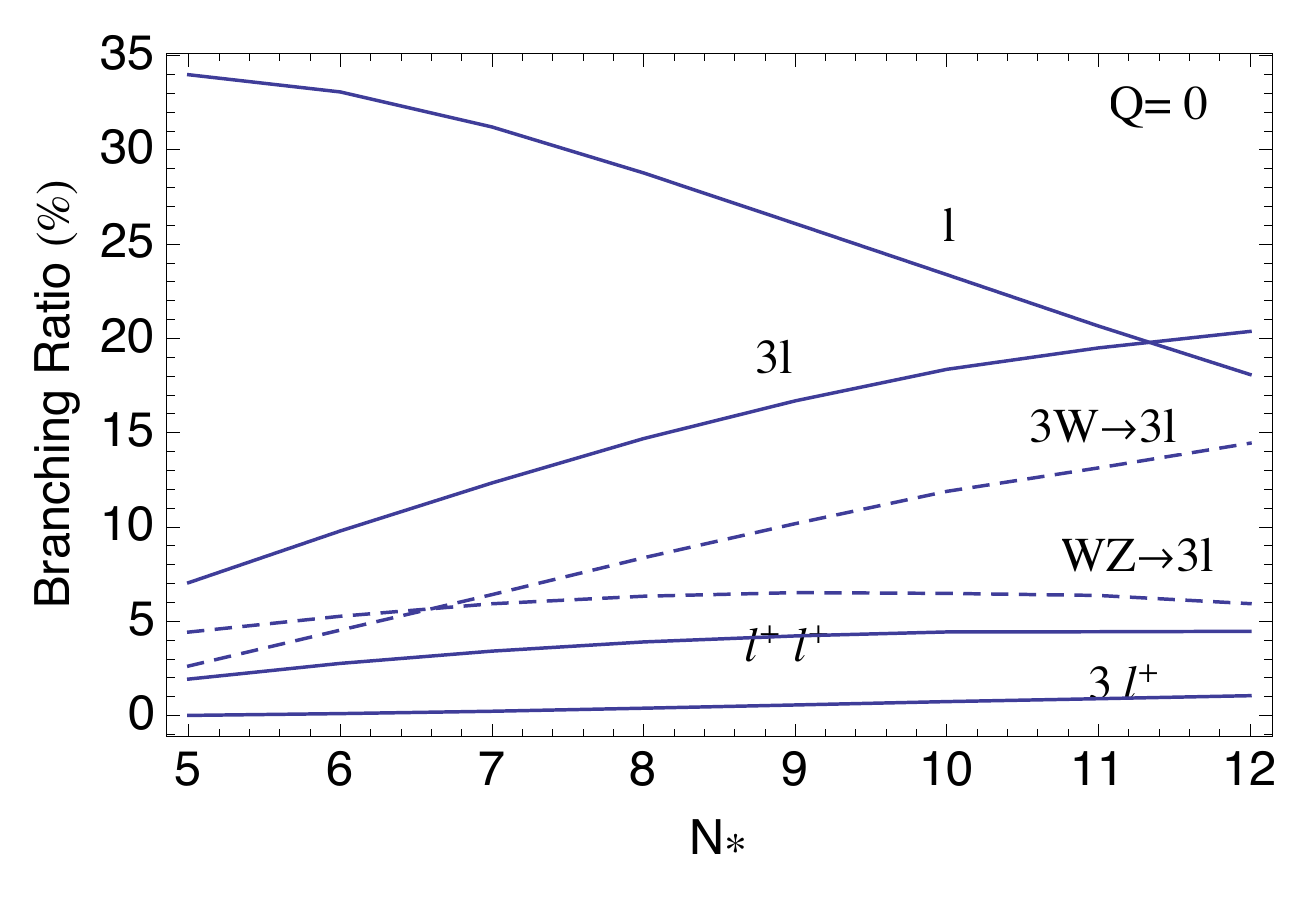} &
\includegraphics[width=0.47\columnwidth]{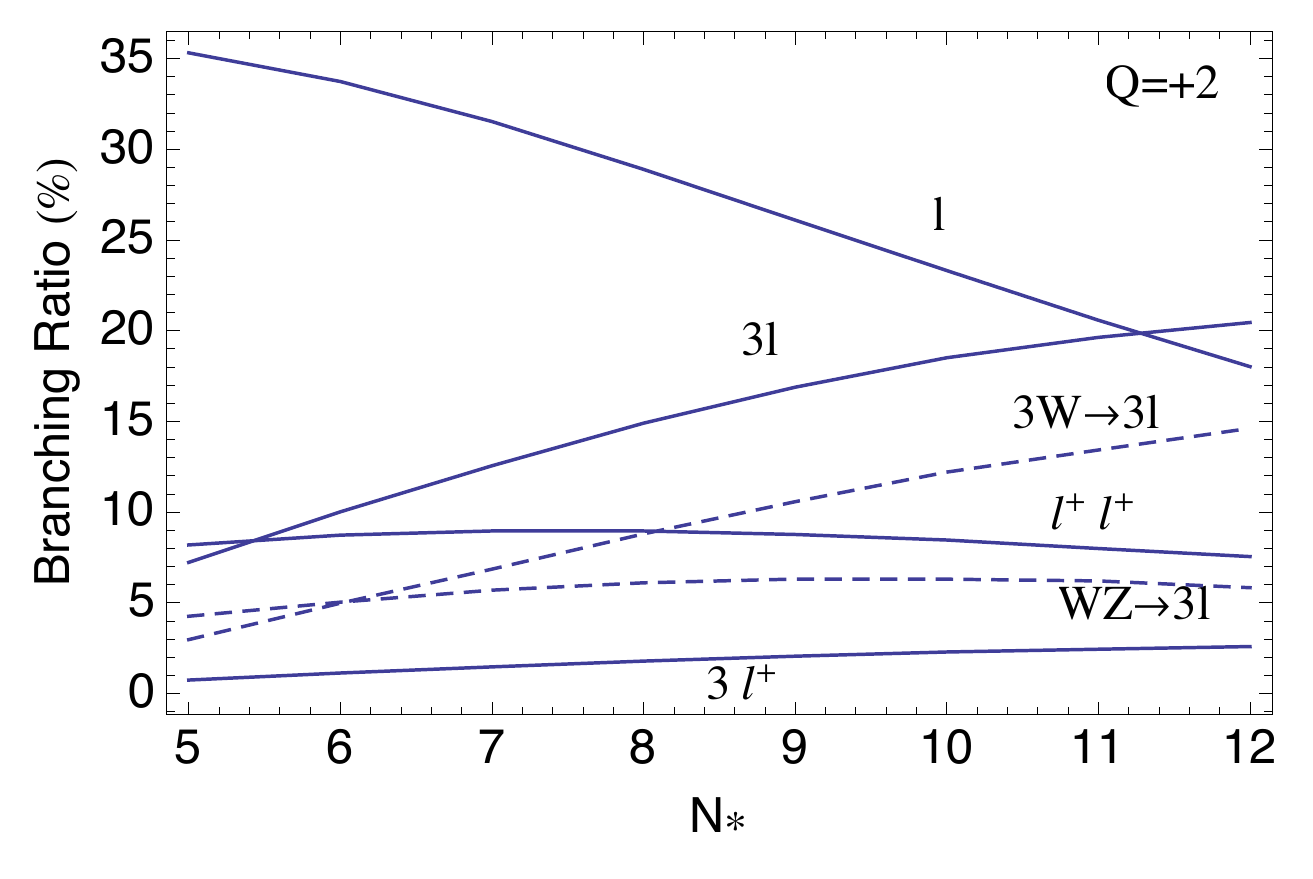} \\
\includegraphics[width=0.47\columnwidth]{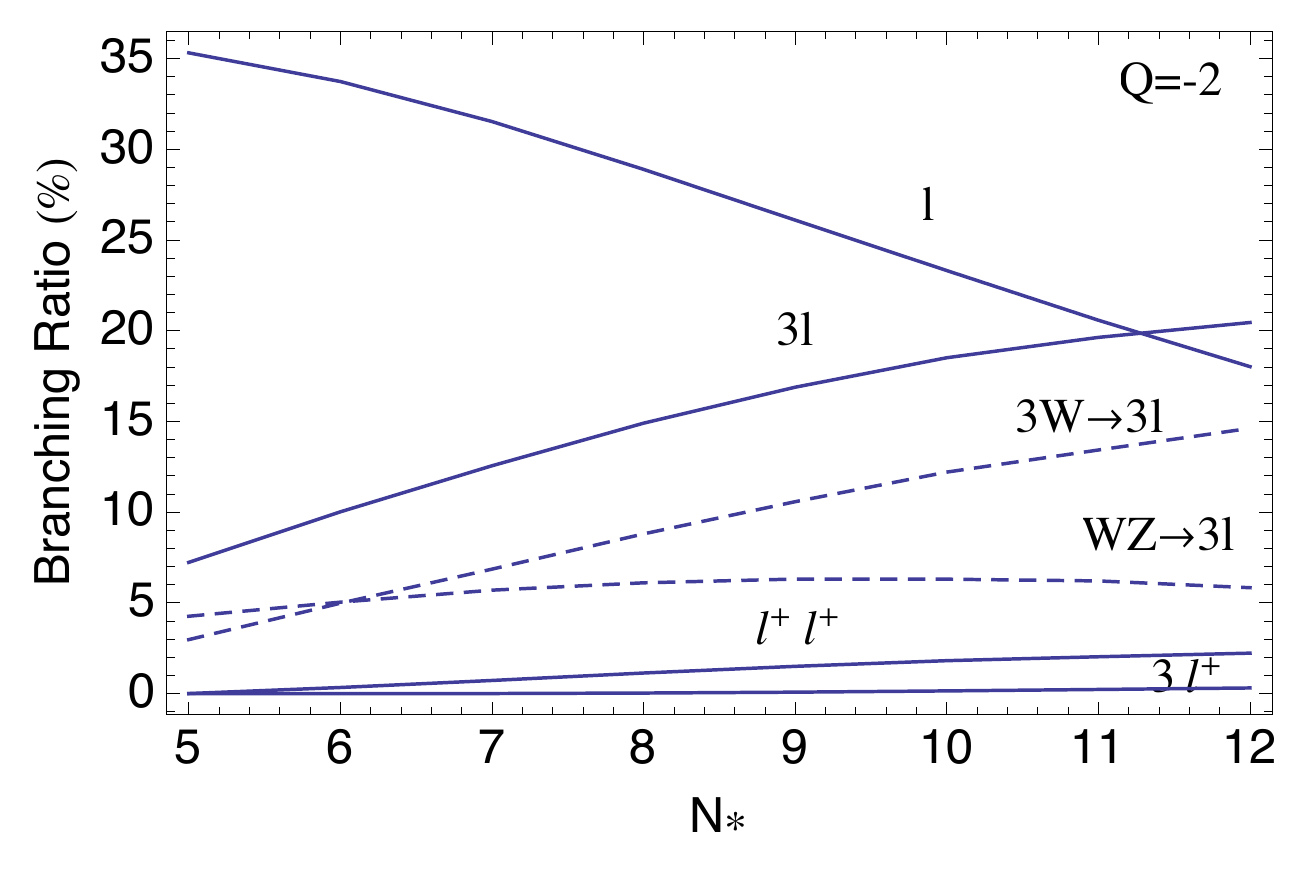}&
\includegraphics[width=0.47\columnwidth]{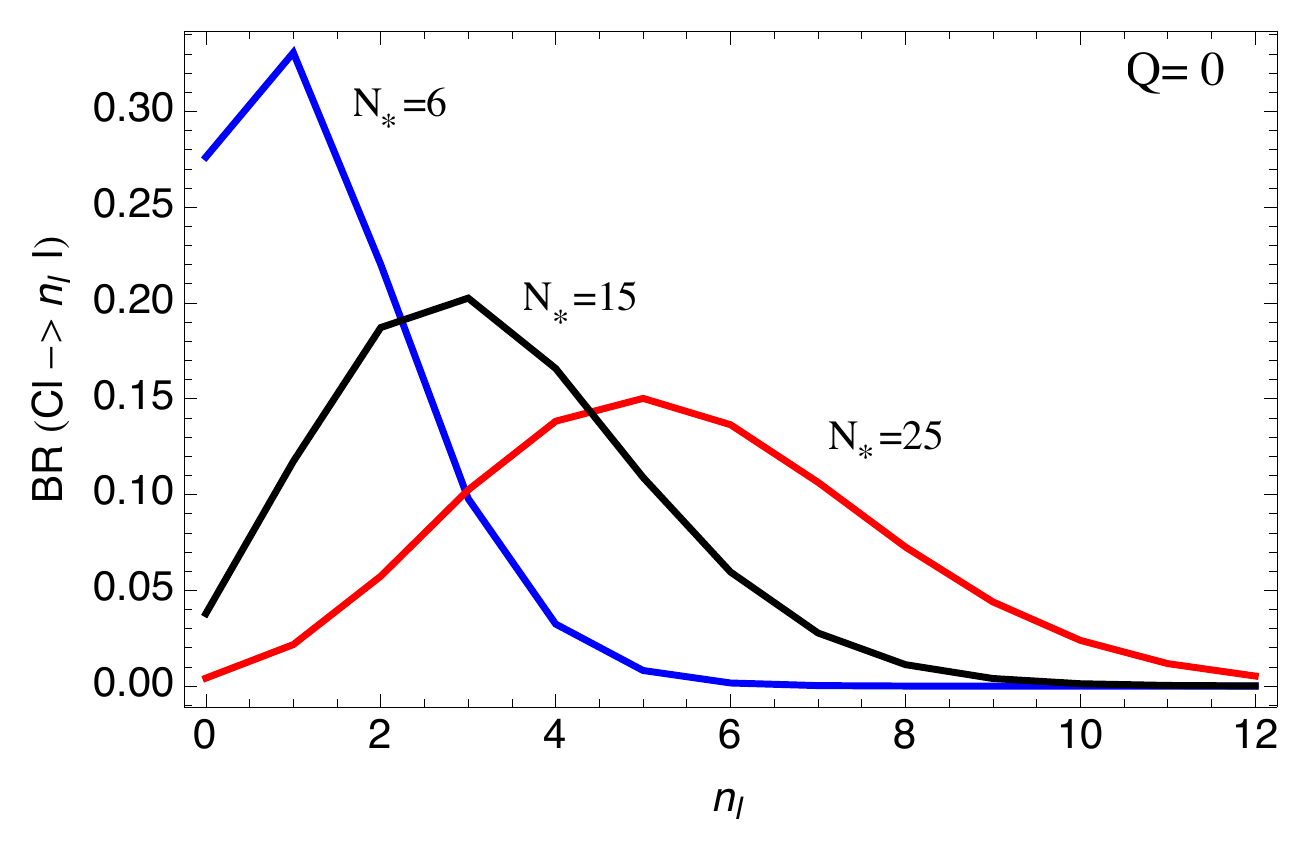} 
\end{tabular}
\caption{In the model of classicalization of  longitudinal $W$s and $Z$s we  show the branching fractions for a neutral classicalon (top left) and for classicalons with electric charge, $Q=\pm2$ (top right/bottom left). In the figure on the bottom right, we show the classicalon branching ratio to  $n_l$ leptons for a neutral classicalon. In the decay channels shown above we require exactly (and not at least)  the number of leptons mentioned. }
\label{nfig1}
\end{figure}

It is important to note that classicalons must have a discrete mass spectrum as was shown in Ref.~\cite{Dvali:2011nh}. The allowed masses are precisely the points marked in Fig.~\ref{nfig} (left), that is masses that give an integer value for $N_*$. At intermediate energies in between two allowed masses,  a classicalon with a lower mass would be formed along with some SM particle(s)~\cite{Dvali:2011nh} that carries the rest of the energy and momentum. We will assume in our analysis that at these intermediate energies the closest classicalon with a lower mass, say $M_{N_*}$, is formed with the cross-section  $\pi r_*^2(M_{N_*})$. As the spacing between the masses that we have found is greater than the $Z$-mass, the additional SM particle emitted can even  be a $W/Z$ boson. This would mean that  we may be able to get $(N_*+1)$ $W/Z$s in the final state even at energies lower than $M_{N_*+1}$. We will avoid this complication as   by ignoring this effect,  which enhances the signal, we are only being conservative.

\subsubsection{Branching ratios}
\label{brs}
In order to derive the classicalon branching ratio to a particular number of $W^+$, $W^-$ and $Z$s, we will assume that  a classicalon  decays democratically and randomly to the three Goldstone components $\pi_+$, $\pi_-$ and $\pi_3$, the only constraint being electrical charge conservation. By the Goldstone boson equivalence principle, we will thus get in the unitary gauge a number of longitudinal $W^+$, $W^-$ and $Z$s equal to the number of $\pi_+$, $\pi_-$ and $\pi_3$s in the final state. Thus the unnormalized probability of a particular $N_*$-particle classicalon composition with number of $W^+$  bosons equal to $N_{W^+}$,  number of  $W^-$  bosons equal to  $N_{W^-}$ and number of  $Z$  bosons equal to  $N_{Z}$,  must be proportional to the number of possible ways of exchanging the identical particles amongst themselves to give the same final state, that is,  
\beq
P'(N_{W^+},N_{W^-},N_{Z})=\frac{N_*!}{N_{W^+}!N_{W^-}!N_{Z}!}.
\label{prob1}
\eeq
 So for instance for a neutral classicalon with energy and radius such that we get $N_*=5$ using Eqs.(\ref{consm1}) and (\ref{consm2}), the possible compositions are: $ ZW^+ W^- W^+ W^-$,  $Z ZZ W^+ W^-$ and $ZZZZZ$.  Computing probabilities as described above we get for these different possibilities,
\bea
P'(Z W^+ W^- W^+ W^-) &=&\frac{5!}{2!2!1!} \nonumber\\
P'(Z ZZ W^+ W^-) &=&\frac{5!}{1!1!3!} \nonumber\\
P'(ZZZZZ) &=& \frac{5!}{0!0!5!}.
\label{prob}
\eea
Finally these probabilities must be normalized,
\beq
P(N_{W^+},N_{W^-},N_{Z})=\frac{P'(N_{W^+},N_{W^-},N_{Z})}{\sum P'(N_{W^+},N_{W^-},N_{Z})}.
\label{prob2}
\eeq
The sum in the above equation runs over all $N_{W^+}, N_{W^-}$ and $N_{Z}$ respecting $N_{W^+}+N_{W^-}+N_{Z}=N_*$ and $N_{W^+}-N_{W^-}=Q$, $Q$ being the electric charge of the classicalon. 

To find the branching fraction to leptons, jets and missing energy that the $Ws/Zs$ decay to, we need to consider still more combinatoric possibilities. We discuss this in detail in Appendix C, where   we provide expressions for the branching ratio to  final states with varying number of leptons.

In Fig.~\ref{nfig1} we show the branching ratio of classicalons with charge, $+2,0$ and $-2$. Note that the branching ratio for decay channels with higher number of leptons rise with $N_*$ whereas the branching ratio of the single lepton channel falls. This is so because classicalons with higher $N_*$ decay to more leptons. This is clear from Fig.~\ref{nfig1} (bottom right) where we show the classicalon branching ratio to  $n_l$ leptons.  We see that an $N_*$-particle classicalon decays with maximum branching ratio to $n_l \sim N_*/5$ leptons. Note that the branching ratios in Fig.~\ref{nfig1}  have been computed at the theoretical level and do not include any experimental effects.

\subsubsection{Signals at the LHC}
 \begin{figure}[t]
\centering
\hspace{-0 in}
\begin{tabular}{cc}

\includegraphics[width=0.6\columnwidth]{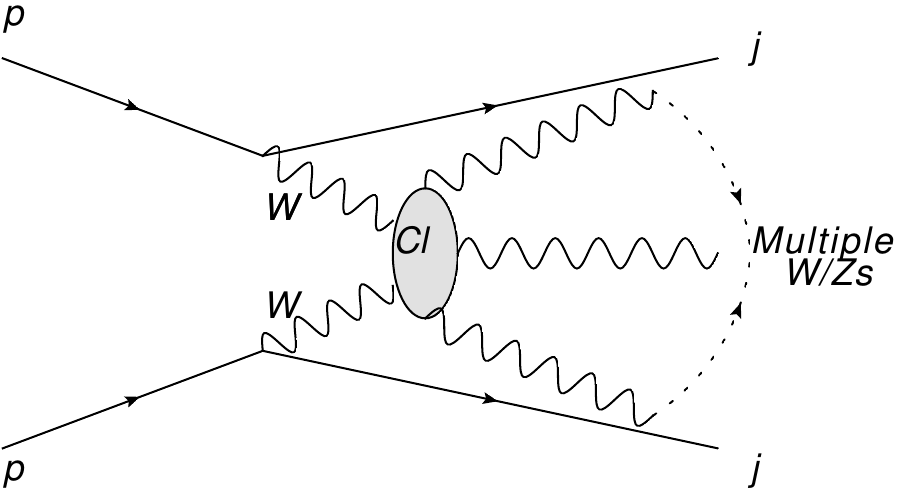}
\end{tabular}
\caption{Production of a classicalon by weak boson fusion  in the model with goldstone classicalization.}
\label{feyndig1}
\end{figure}
\begin{sidewaystable}[ph!]
\centering
\begin{tabular}{|c|c|c|c|c|c|c|c|c|}
\hline
\multirow{2}{*}{$M$(GeV)}&\multirow{2}{*}{$N_*$}&\multicolumn{3}{|r|}{Total Cross-section (fb)}&\multicolumn{4}{|c|}{Signal Cross-section}\\

&&Q=+2&Q=0&Q=-2&$l$(fb)&$l^+l^+$(fb)&$3l (3W\to 3l, WZ\to 3l)$ (fb)&$3l^+$(fb)\\
\hline
\multicolumn{9}{|l|}{$M_*=246$ GeV, LHC energy = 7 TeV}\\
\hline
$830$&$6$&8.3&7.6&1.7&5.9 [2.7]&0.9 [0.5]&1.7 (0.8, 0.9) [1.0 (0.5, 0.5)]&0.1 [0.06]\\
$958$&$7$&4.7&4.1&0.9&3.0 [1.2]&0.6 [0.3]&1.2 (0.6, 0.6) [0.6 (0.3,0.3)]&0.07 [0.04]\\
$1086$&$8$&2.7& 2.3& 0.5&1.6&0.3&0.8 (0.5, 0.3)&0.05\\
$1213$&$9$&1.6& 1.3& 0.2&0.8&0.2&0.5 (0.3, 0.2)&0.04\\
$1340$&$10$&1.0& 0.7& 0.1&0.4&0.1&0.3 (0.2, 0.1)&0.03\\
$1467$&$11$&0.6& 0.4& 0.07&0.2&0.07&0.2 (0.1, 0.07)&0.02\\
$1594$&$12$&0.4& 0.3& 0.04&0.1&0.04&0.2(0.1, 0.05)&0.01\\
\hline
\multicolumn{9}{|l|}{$M_*=246$ GeV, LHC energy = 14 TeV}\\
\hline
$830$&$6$&74 &84&23&60 [28]&8.9 [4.8]&18 (8.6, 9.3) [11 (5.5, 5.0)]&0.9 [0.6]\\
$958$&$7$&50&55&15&38 [15]&6.4 [3.0]&15 (8.0, 6.9) [7.5 (4.3, 3.2)]&0.9 [0.5]\\
$1086$&$8$&33&35&9.2&22&4.4&11 (6.6, 4.8)&0.7\\
$1213$&$9$&23&24&5.9&14&3.1&8.9 (5.5, 3.4)&0.6\\
$1340$&$10$&16& 16& 3.9&8.4&1.9&6.6 (4.3, 2.3)&0.5\\
$1467$&$11$&12& 11& 2.7&5.3&1.5&5.0 (3.4, 1.6)&0.4\\
$1594$&$12$&8.8& 8.3& 1.9&3.4&1.1&3.9 (2.8, 1.1)&0.3\\
\hline
\multicolumn{9}{|l|}{$M_*=600$ GeV, LHC energy =14 TeV}\\
\hline
$1160$&$6$&3.1& 3.2& 0.8&2.4 [1.1]&0.6 [0.2]& 0.7 (0.3, 0.4) [0.4 (0.2, 0.2)]& 0.04 [0.02]\\
$1325$&$7$&2.0& 1.9&0.5&1.4 [0.5]&0.2 [0.1]&0.5 (0.3, 0.2) [0.3 (0.2, 0.1)]&0.03 [0.02]\\
$1490$&$8$&1.3& 1.2& 0.3&0.8&0.2& 0.4 (0.2, 0.2)& 0.03\\
$1655$&$9$&0.9& 0.8& 0.2 &0.5& 0.1& 0.3 (0.2, 0.1)& 0.02\\
$1820$&$10$&0.6& 0.5& 0.1&0.3& 0.07& 0.2 (0.1, 0.07)& 0.02\\
$1980$&$11$&0.4& 0.4& 0.08&0.2& 0.05& 0.2 (0.1, 0.06)& 0.01 \\
$2140$&$12$&0.3& 0.3& 0.05&0.1& 0.04& 0.1 (0.09, 0.04)& 0.01\\
\hline
\multicolumn{9}{|l|}{$M_*=1$ TeV, LHC energy =14 TeV}\\
\hline
$1580$&$6$&0.3& 0.3& 0.07&0.2 [0.1]&0.03 [0.02]&0.07 (0.03,0.03) [0.04 (0.02,0.02)]& -\\
$1795$&$7$&0.2& 0.2& 0.03&0.1 [0.05]&0.02 [0.01]&0.05 (0.03,0.02) [0.03 (0.02,0.01)]& -\\
$2005$&$8$&0.1& 0.1&0.02&0.06&0.01&0.03 (0.02,0.01)& -\\
$1215$&$9$&0.1& 0.1& 0.01 &0.05&0.01&0.03 (0.02, 0.01)& -\\
$2420$&$10$&0.05& 0.04&0.01&0.02&0.01&0.02 (0.01, 0.01)& -\\
$2620$&$11$&0.03& 0.03& 0.01&0.01&-&0.01 (0.01, -)& -\\
$2820$&$12$&0.02& 0.02& 0.01&-&-&0.01 (0.01, -)&-\\
\hline
\end{tabular}
\caption{ Cross-section for classicalon production by weak boson fusion in the model with goldstone classicalization. We give the total cross-section as well as the cross-section in the different channels. The number of leptons mentioned in each channel is the exact number of leptons in the final state. The values in the square brackets are the cross-section values assuming no invisible $Z$ decays  and no $W$ decays to hadronically decaying $\tau$s, which ensures the maximum number of partonic jets. Note that no effect of showering, hadronization, experimental cuts or detector acceptances has been included here. For a discussion of these, see the text.}
\label{bigt}
\end{sidewaystable}

At the LHC these classicalons can be produced in the weak boson fusion (WBF) process, $pp \to jj(W_L W_L \to Cl$) (see Fig.~\ref{feyndig1}). To  compute the cross section for their production we use the effective $W$ approximation. In this approximation the luminosity of longitudinal $W$ bosons  is  given by~\cite{Dawson:1984gx},
\beq
\frac{d L}{d \tau}=\left(\frac{g^2}{16 \pi^2}\right)^2\frac{1}{\tau}[(1+\tau)\ln(1/\tau)+2(\tau -1)]
\eeq
where $\tau = \hat{s}/s_q$ is the ratio of the squared center of mass energy of the $W$-pair, $\hat {s}$, to the squared center of mass energy of the initial quarks, $s_q$. The cross-section for production of an $N_*$-particle classicalon is found by convoluting the geometric cross-section with this luminosity function and the  parton density functions as follows,
\beq
\sigma_N =\sum_{ij}\int_{M^2_{N_*}/s}^{M^2_{N_*+1}/s} d \tau~ \pi r_*^2(M_{N_*}) \int_{\tau}^{1} \frac{d \tau'}{\tau'}\int_{\tau'}^{1} \frac{dx}{x} f_i(x,q^2) f_j (\tau'/x,q^2)\frac{dL}{d \xi}
\label{crosssec}
\eeq
where now $\tau = \hat{s}/s$,  $s$ being the proton-proton center of mass energy squared, $\tau'=s_q/s$ and  $\xi =\tau/\tau'$. We have taken the factorization scale $q^2= M_W^2$.  As we stated already, we have assumed that for energies $M_{N_*}<\hat{s}<M_{N_*+1}$, an $N_*$-particle classicalon is formed along with other SM particles  with a cross-section $\pi r_*^2(M_N)$. For our computations we have used the MSTW parton density functions (PDF)~Ê\cite{Martin:2009iq}. In the  summation above both  $i$ and $j$  run over all positively charged quarks for $W^+ W^+$ fusion which leads to production of classicalons with charge $+2$, and  run over all negatively charged quarks for $W^- W^-$ fusion which leads to production of classicalons with charge $-2$. For production of neutral clasicalons from $W^+W^-$ fusion $i$ and $j$ run over quarks with opposite electric charge. As we are considering only $W^{\pm}$ in the initial state, the classicalons produced can have charge only $-2, 0$ and $+2$. The contribution of initial states with a $Z$ boson has been neglected here as the $Z$ boson  luminosity is much smaller compared to the $W$ boson luminosity. For instance, the  $ZZ$ luminosity is about an order of magnitude smaller than the $W^+W^-$ luminosity~\cite{Dawson:1984gx}. 

 The final states that would be seen in colliders are leptons plus multijets and missing energy. We will provide cross-sections for the final states, $l +\Emisst+ jets$, $l^+l^+ +\Emisst+jets$, $3l+\Emisst+jets$ and   $3l^+ +\Emisst+jets$ where $l$ can be an electron, muon or leptonically decaying $\tau$ and we consider hadronically decaying $\tau$s as jets. In the final states above we require exactly (and not at least)  the number of leptons mentioned. While the $l +\Emisst+ jets$ channel would be the discovery mode with the highest cross-section a simultaneous observation of a signal in the other more striking channels,  $l^+l^+ +\Emisst+jets$, $3l+\Emisst+jets$ and   $3l^+ +\Emisst+jets$, would provide confirmation that the phenomenon is indeed classicalization.  The fact that missing energy must be present in these channels is an important difference from the black hole case where the probability of  neutrino emission is small ($< 5 \%$) and one can have final states with leptons and jets but no missing energy (this is the final state discussed in Ref.~\cite{Dimopoulos:2001hw} for instance).  As we said earlier, when we go to higher $N_*$ values channels with even more leptons will become important. The production cross-section for classicalons, however, decreases as $N_*$  increases because of the falling longitudinal $W$ luminosity in Eq.(\ref{crosssec}).\footnote{As we will soon see, another issue  for channels with greater number of leptons is that there is a greater reduction   in cross-section for these channels when experimental requirements like lepton isolation are taken into account. } We will, therefore, not study  channels with  larger number of leptons.

The production cross-section for classicalons  is given in Table~\ref{bigt}  for $M_*=246$ GeV, $M_*=600$ GeV and $M_*=1$ TeV at LHC energies 7 and 14 TeV. We have provided contributions only for $N_* \geq 6$.\footnote{The energy regime close to the classicalization scale that we have not considered, that is  $\sqrt{\hat{s}}\sim M_*$ and  $N < 6$, would phenomenologically resemble strong electroweak symmetry breaking (EWSB) theories like technicolor with the appearance of quantum resonances at this scale.  Final states with as many as five final $W/Z$s have already been mentioned in the literature as signatures for strong EWSB~Ê\cite{Evans:2009ga}.} Using the branching ratios evaluated in the previous section, we give in Table~\ref{bigt}  the cross-sections of the four channels, $l $, $l^+l^+$, $3l$ and   $3l^+$,  that we are interested in. For the $3l$ channel there are two different ways in which three leptons can be produced, from the decay of three $W$s or from the decay of a $W$ and $Z$. In Table~\ref{bigt} we provide the individual contribution from both these channels as these two modes can be experimentally distinguished by checking if a lepton pair reconstructs the $Z$-mass. Also, the number of partonic jets is higher for the $WZ\to 3l$ mode than the  $3W\to 3l$ mode. The number of partonic jets in an event is maximum if all the  $Z$s that do not decay leptonically, decay hadronically (and not invisibly) and all the $W$s  that do not decay leptonically, decay to quarks pairs and not to   $\tau$-jets.  In Table~\ref{bigt} we have given in square brackets for $N_*=6$ and $N_*=7$, the cross-section values assuming  the maximum possible number of jets are produced. As one can see from these values for $N_*=6$ and $N_*=7$ about half of the time the classicalon does decay to the maximum number of jets possible.  

\begin{figure}[t]
\centering
\hspace{-0.2 in}
\begin{tabular}{cc}
\includegraphics[width=0.47\columnwidth]{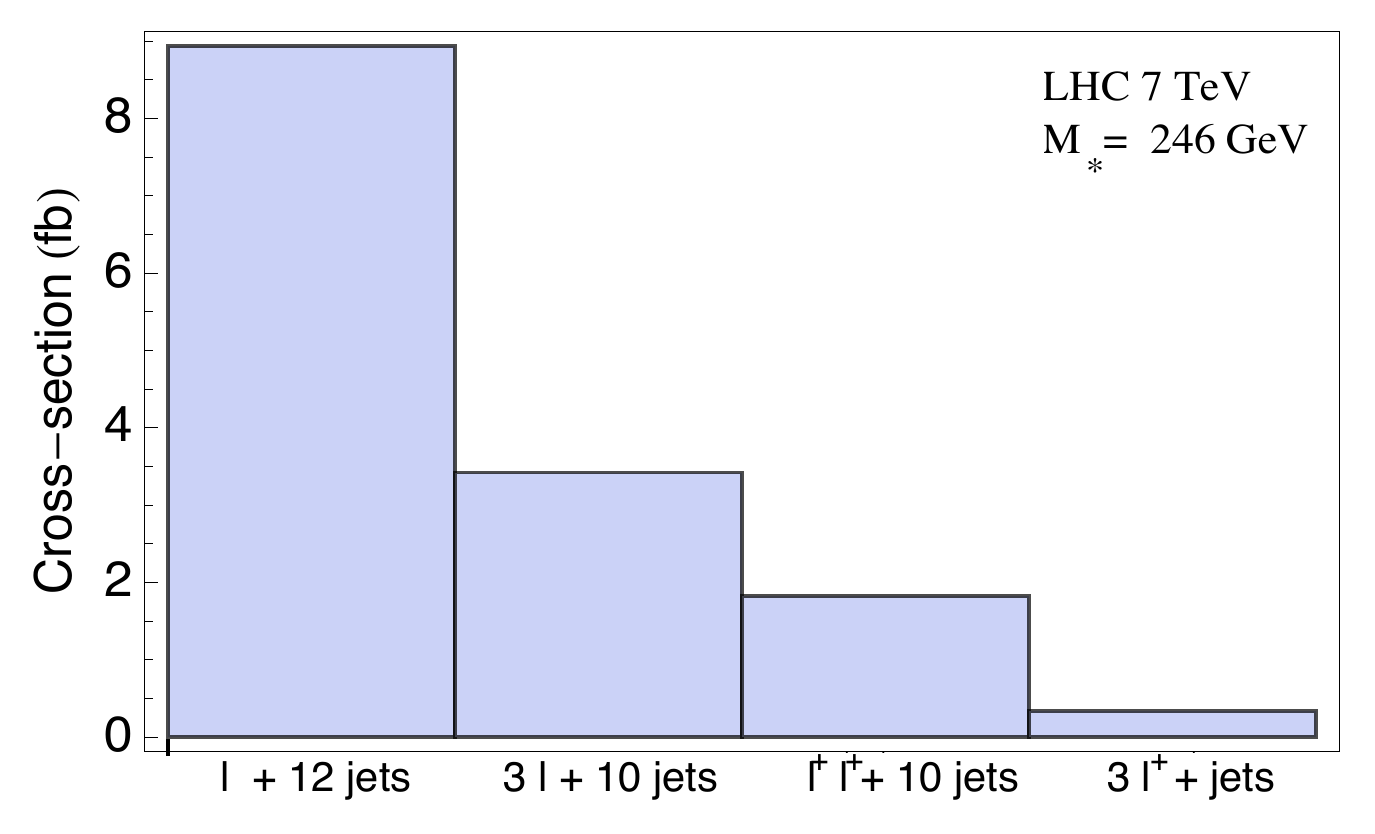} &
\includegraphics[width=0.47\columnwidth]{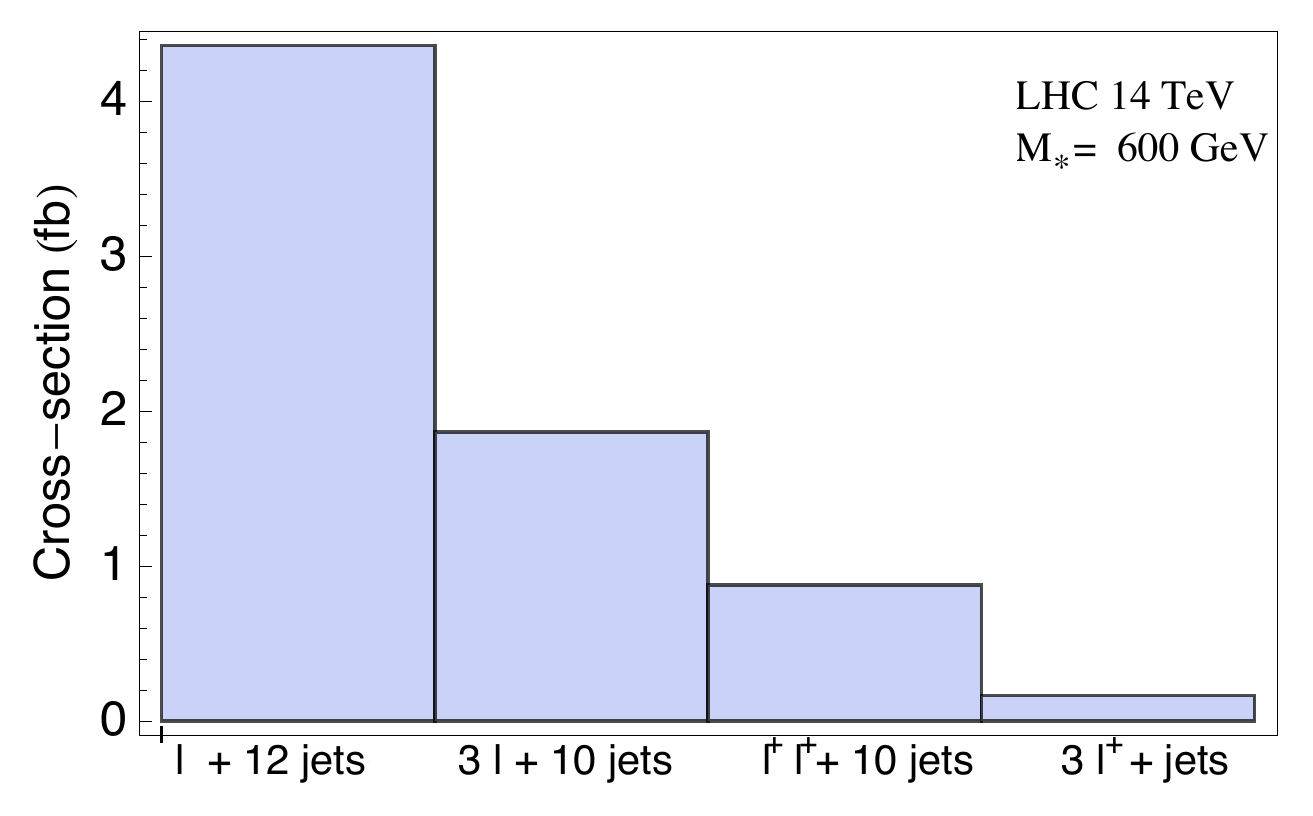} \\
\includegraphics[width=0.47\columnwidth]{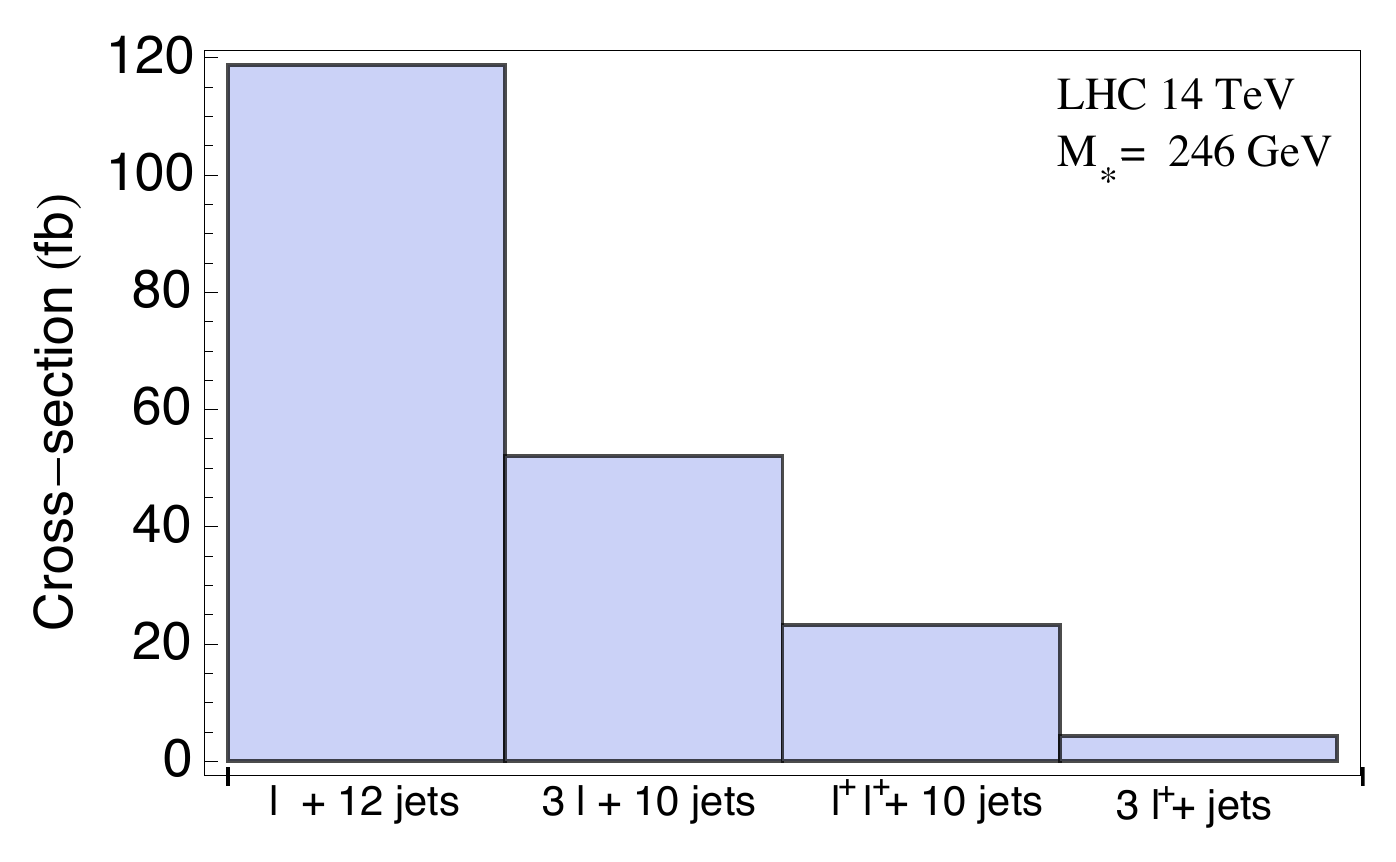} &
\includegraphics[width=0.47\columnwidth]{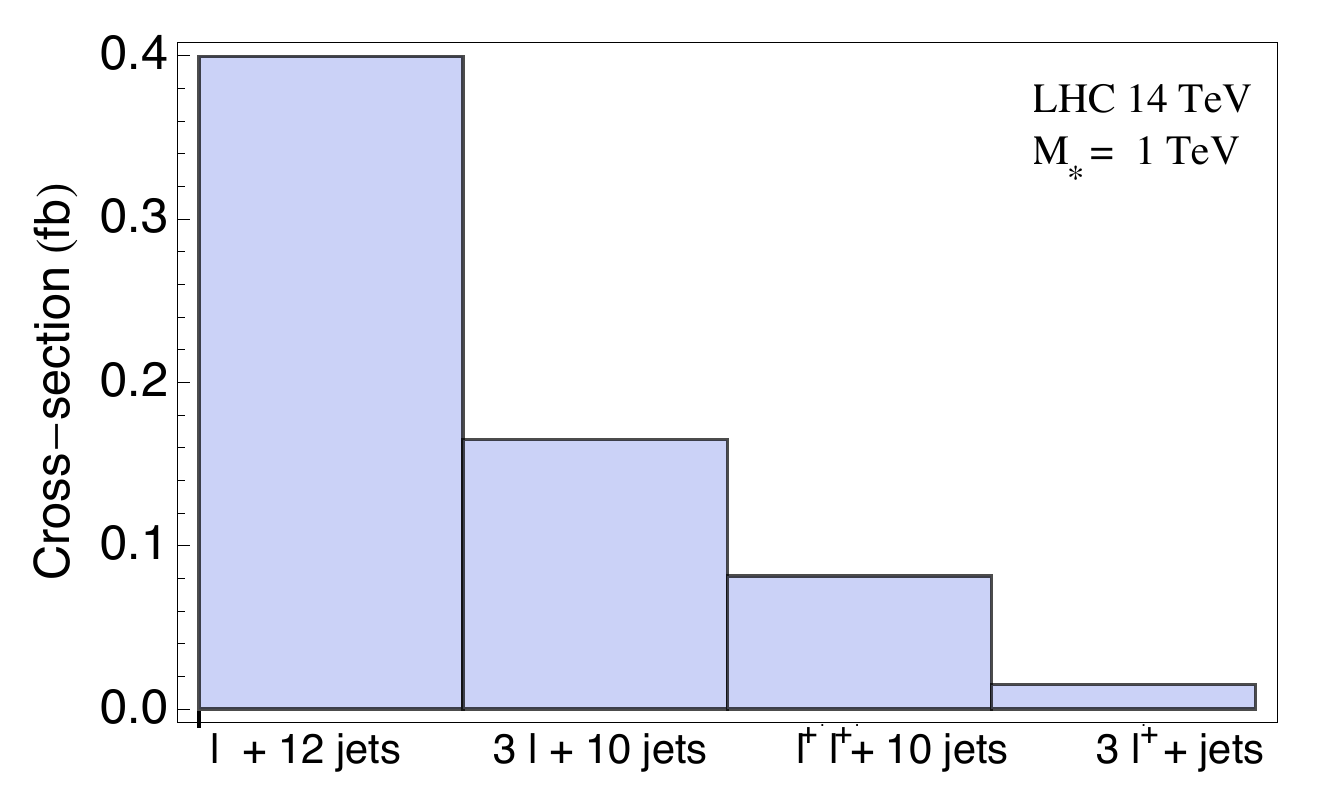}
\end{tabular}
\caption{Cross-section for production of  a lepton plus at least 12 partonic jets, two same sign leptons plus at least 10 partonic jets,  three leptons plus at least 10 partonic jets and for three same sign leptons  from  the decay of classicalons formed by longitudinal $W$s and $Z$s. Missing transverse energy is present in all the cases mentioned above and the number of leptons mentioned in each case is the exact number of leptons in the final state.  Note that the number of jets mentioned above is at the partonic level and no effect of showering, hadronization, experimental cuts or detector acceptances has been included here. For a discussion of these effects see the text.}
\label{bars}
\end{figure}

The number of jets produced is very large and this  ensures that the background is negligible. Including the two forward jets produced in the WBF process, for $N_*=6 (10)$  as many as 12 (20) partonic jets in the single lepton channel, 10 (18) partonic jets in the $l^+l^+$ and $3l$ channels, and 8 (16) partonic jets in the   $3l^+$ channel, can be produced. In Fig.~\ref{bars} we add up contribution from classicalons with $N_* \geq 6$ and show the  cross-section for $l$ plus at least 12 partonic jets, $l^+l^+$ plus at least 10 partonic jets,  $3l$ plus at least 10 partonic jets and the cross-section for the  $3l^+ + \Emisst+jets$ channel.\footnote{ For evaluation of the exclusive contribution to the  cross-section of $l +  12~jets$ from $N_*=6$ classicalon decays, $l^+l^+ + 10~jets$ from $N_*=6$ classicalon decays, $(WZ \to 3l)+10~jets$ from $N_*=6$ classicalon decays and $(3W \to 3l)+10~jets$ from $N_*=7$ classicalon decays we have to use the  cross-section values allowing no invisible decays of the $Z$ boson and no tau decays of $W$ bosons, as such decays would lead to fewer  jets than the required number. Decay of two or more $Z$s invisibly   is relatively unlikely and has been  ignored here.} We do not require a minimum number of jets in the last case as the background is absent even without this requirement. Whereas the $3l^+$  channel is virtually background free (the background cross-section is of the order of  0.01 fb at 14 TeV LHC~\cite{Mukhopadhyaya:2010qf}) the other channels also have negligible background if we require so many jets. The $l +\Emisst +jets$ background gets its major contribution from the $t \bar{t}+ jets$ production and for more than 10 jets the background, with appropriate cuts,  is negligible~\cite{Lisanti:2011tm}. The $l^+l^+  +\Emisst + jets$  background has been discussed in detail in Ref.~\cite{Contino:2008hi}
and Ref.~\cite{Dissertori:2010ug} and again cuts can be applied to reduce  this background  to a negligible value for high jet multiplicities (8 or more jets). As all the major  SM processes that contribute $3l +\Emisst +jets$ background, like the $WZ +jets$ process, would also contribute to the single lepton channel, if the $l +\Emisst +jets$ background is negligible, this background can also be neglected at high jet multiplicities.

It should be noted that the cross-section values in Table~\ref{bigt} and Fig.~\ref{bars} do not include any effect of parton showering, experimental cuts or detector acceptances. Let us discuss the important experimental effects not taken into account here. The experimental cut that is expected to have a substantial effect in the presence of so many jets  is the requirement for lepton isolation. For instance if we consider 15 partonic  jets having a cone radius $\Delta R =\sqrt{\Delta \eta^2+ \Delta \phi^2}=0.4$, we can roughly estimate the fraction of times an isotropically emitted lepton would remain isolated by finding the fraction of area of in $\eta- \phi$ space that is still unoccupied by the jets assuming conservatively that the jets do not overlap. To take into account the fact that the leptons and jets are produced centrally we limit their $\eta$-range to $-1.5 < \eta < 1.5$, which gives a total allowed area $\Delta \eta \Delta \phi = 6 \pi$. This  estimate gives us  about  60$\%$ probability that a lepton would be isolated for 15 non-overlapping jets. It should also be kept in mind that that the lepton identification rate is about 90$\%$~\cite{tdr}. Thus, this estimate tells us,  due to the requirement of all the leptons being isolated and getting  identified, the cross-section would be reduced to about 54$\%$ of the theoretical value in the $l +\Emisst + jets$ channel, to about $29\%$ of the theoretical value in the $l^+ l^+  +\Emisst + jets$ channel and to about $16\%$  of the theoretical value  in the channels with three leptons. At the same time, the  $3l$ ( $l^+ l^+$) channel would contribute about 35$\%$(50$\%$) of the time to the single lepton channel when not all but only two (one) of the leptons are lost due to lepton isolation/identification requirements. A similar contribution from the $3l$ channel to the $l^+l^+$ channel would be relatively  small. As the leptons are produced isotropically, $p_T$ and $\eta$ cuts are not expected to have a big effect.  Now we come to the experimental cuts related to the jets. A limitation of our analysis that it has been carried out at the partonic level only. Whereas the number of jets would increase from the number at the partonic level because of  parton showering,  other experimental effects like $p_T$ and $\eta$ cuts and most importantly the  jet isolation cut requiring a minimum $\Delta R$ separation between any two jets, would decrease the number of jets from the partonic level. The $\Delta R$ cut is important because if the number of jets is  very large and it is  likely for two or more partonic jets to  merge thus reducing the number of jets experimentally observed. The number of jets produced in a classicalon decay is so large, however, that even after a possible reduction due to the above factors we would expect many jets.  Finally, an experimental cut requiring a minimum missing transverse energy should not reduce the signal cross-section appreciably.

Keeping these issues in mind we see from Fig.~\ref{bars} that whereas for $M_*= 246$ GeV, classicalization should be seen in the $l+\Emisst+jets$ channel (with hints seen in the other channels also) in the present run of the  LHC  with about 10 fb$^{-1}$ data, a thorough confirmation with observation in all the channels would require data at 14 TeV. On the other hand for $M_*= 600$ GeV about 10 fb$^{-1}$ data at 14 TeV would be needed for both discovery and confirmation in  the different channels. The cross-section for $M_*=1$ TeV is about ten times smaller than that for $M_*=600$ GeV and   this is the maximum classicalization scale that can be probed with about 100 fb$^{-1}$ LHC data at 14 TeV.

Another important measurement would be the dependence  of $N_*$ on the total energy of the decay products shown in Fig.~\ref{nfig} (left).  It is theoretically equivalent to measure the average energy of a lepton/partonic jet, $M/(2N_*)$, in the classicalon rest frame\footnote{The typical energy measured in the lab frame would not be so different from the typical energy in the classicalon rest frame  because we expect, as is  the case in  black hole production~\cite{Giddings:2001bu},  that the  classicalons produced would not be highly boosted.} as a function of the total invariant mass. Experimentally, however, the average energy of a lepton/partonic jet is a more tractable quantity than the total multiplicity as it is not affected even if there is missing energy. We plot the average energy of a lepton/partonic jet as a function of the mass  in Fig.~\ref{nfig} (right).  We see that the average energy decreases very gradually. As far as  leptons are concerned it should be straightforward to measure the typical energy. An interesting feature to be checked  would be that the typical lepton energy should be same in all the different channels $l +\Emisst + jets$, $l^+l^+ \Emisst +jets$, $3l+\Emisst +jets$ and   $3l^+ +\Emisst +jets$. To find the typical energy of a jet in an event as a function of the total energy and confirming that this is same as the typical lepton energy would be  much more complicated. This is again because the energy of a jet at the partonic level is not the same as the final energy measured in the detector. The typical jet energy would decrease due to parton showering and increase if two jets get merged.  Another error in the measurement would come from the fact that two of the jets in the event would be the WBF jets which would not have the typical energy in Fig.~\ref{nfig} (right), but this would not be a large effect because of the large number of jets present.   Simulations including parton showers, hadronization and jet algorithms are needed in order to trace back the energy at the partonic level from the final energy measured in the detectors. 

\subsection{Higgs as the classicalizer}

The second  application of classicalization we want to consider is a model where  the classicalizing field is the Higgs itself and the classicalons (called Higssions in this case) are configurations of the Higgs field. The motivation for this model comes from the hierarchy problem. Indeed, the radiative corrections to the Higgs mass in this model are screened by the classization scale itself and not by the highest possible UV scale. In other words, the loop contributions to the Higgs mass get classicalized and cut-off at the classicalization scale $M_*$. As the biggest contribution to the Higgs mass comes from the top, the top loop must get classicalized at the lowest scale, that is for the Higgs mass to be natural we must have,
\beq
\frac{y_t}{16 \pi^2}M_*^2\sim m_h^2
\eeq
where $M_*$ is the classicalization scale. This gives the condition $M_* \lesssim 4 \pi m_h$. We will consider the case where only the right handed top has a classicalizing interaction of the form,
 \beq
\frac{\kappa}{M_*^2}(H^\dagger H) \bar{t}_R {\not \! \! \partial}_\mu t_R .
\label{tr}
\eeq
 It is reasonable to consider the possibility of a universal classicalization scale for all SM particles,  in which case Higgsions would be produced at low scales directly from the light quarks. This scenario, however, would be far more constrained by existing flavor and  LHC data. Here we will consider the minimal case required for naturalness with only the right handed top having a low scale classicalizing interaction. In this case  the radius of the classicalon is given by the expression~\cite{Dvali:2010jz},
\bea
r_* \sim \frac{\kappa v M}{M_*^{3}} ~~~~~{\rm for~ \kappa>0}\\
r_* \sim \frac{\kappa M}{M_*^{2}} ~~~~~{\rm for~ \kappa<0} 
\eea
where $v$ is the Higgs VEV.  Again the above relationships is valid only until $r_*$ reaches the compton wavelength of the Higgs, $1/m_h$ and beyond this point the radius freezes at the value $1/m_h$~\cite{Dvali:2010jz}. Again we will absorb any numerical coefficient  present in the above expressions for $r_*$  and also the numerical value of  $\kappa$   in a redefinition of $M_*$ to obtain,
\bea
r_* = \frac{ v M}{M_*^{3}} ~~~~~{\rm for~ \kappa>0}\\
r_* = \frac{M}{M_*^{2}} ~~~~~{\rm for~ \kappa<0.}
 \label{neg}
\eea
Again, experimental constraints due to quantum resonances around $M_*$ are  unfortunately incalculable.
\begin{figure}[t]
\centering
\hspace{-0.2 in}
\includegraphics[width=0.7\columnwidth]{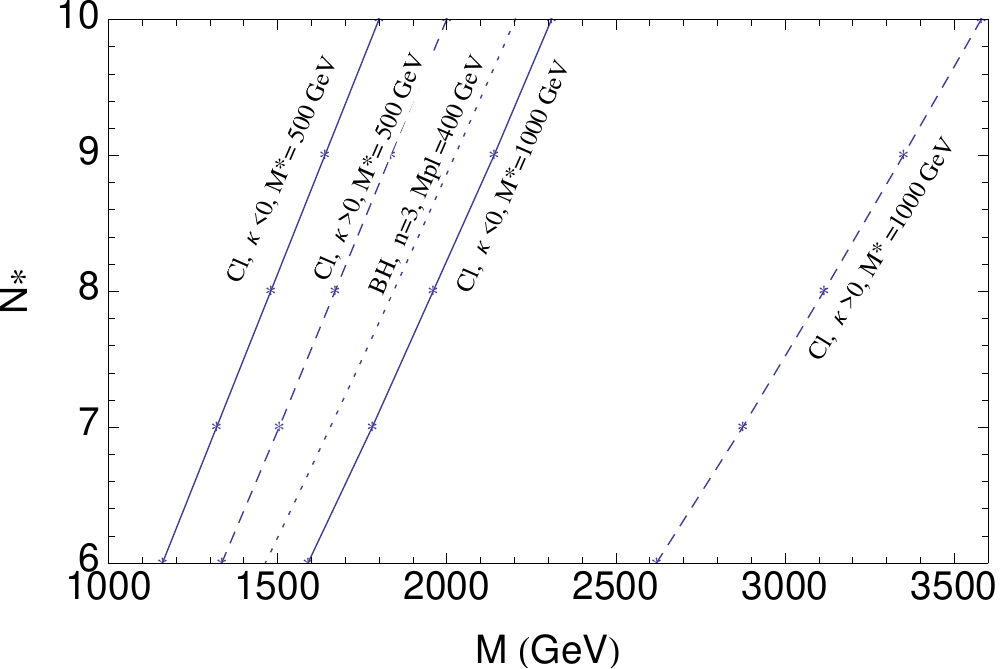} 

\caption{In the model with Higgs as the classicalizer  we plot the number of Higgs bosons produced,  $N_*$, in a classicalon decay as a function of the classicalon mass and compare it with the multiplicity curve for a black hole in $n=3$ extra dimensions  with $M_{pl}=400$ GeV. }
 \label{ntopp}
\end{figure}

\begin{sidewaystable}
\centering
\begin{tabular}{cc|ccc|ccc}
\hline
$N_*$&Branching ratio&\multicolumn{3}{|c|}{Cross-section for $\kappa<0$}&\multicolumn{3}{|c}{Cross-section  for $\kappa>0$}\\
&to $l+$ jets($\%$)&$M$ (GeV)&All channels (fb)&  $l+$ jets(fb)&$M$ (GeV)&All channels (fb)& $l+$ jets(fb)\\
 
\hline
\multicolumn{7}{l}{$M_*= 500$ GeV}\\
\hline
6&29&1160&26(0.5)&7.5(0.2)&1335&8.4(0.08)&2.4(0.03)\\
7&26&1320&17(0.3)&4.4(0.08)&1505&5.4(0.04)&1.4(0.01)\\
8&23&1480&13(0.2)&3.0(0.05)&1670&4.0(0.01)&0.9(-)\\
9&21&1640&8.0(0.1) &1.7(0.02)&1835&3.2(0.01)&0.7(-)\\
10&18&1800&6.0(0.06)&1.1(0.01)&2000&2.1(-)&0.4(-)\\
\multicolumn{2}{c|}{Total cross-section:}&&70(1.2)&18(0.4)&&23(0.1)&5.8(0.04)\\
\hline
\multicolumn{7}{l}{$M_*= 1$ TeV}\\
\hline
6&29&1590&1.0&0.3&2620&0.02&-\\
7&26&1780&0.6&0.2&2875&0.01&-\\
8&23&1960&0.4&0.09&3115&0.01&-\\
9&21&2140&0.2&0.04&3350&-&-\\
10&18&2310&0.2&0.04&3580&-&-\\
\multicolumn{2}{c|}{Total cross-section:}&&2.4&0.7&&0.04&-\\
\hline\\
\end{tabular}
\caption{ Cross-section for classicalon production by top fusion in the model with the Higgs as the classicalizer. We give the total cross-section as well as the cross-section for only the $l+jets$ channel. We have considered the cases $M_*=500$ GeV and $M_*=1$ TeV. The cross-sections  for 7 TeV LHC energy, when not negligible ($<$0.01 fb), are given in parentheses. All other numbers are for 14 TeV LHC energy. For evaluating the branching ratio to the $l+jets$ channel we have used the SM numbers for $m_H =130$ GeV,  that is, $BR(H \to q\bar{q})= 0.55$, $BR(H \to gg)=0.06$, $BR(H \to \tau \tau)=0.05$, $BR(H \to WW)=0.29$ and $BR(H \to ZZ)=0.04$. We have considered hadronically decaying $\tau$s to be jets and leptonically decaying $\tau$s to be leptons. Note that no effect of showering, hadronization, experimental cuts or detector acceptances has been included here. For a discussion of these, see the text.}
\label{sens}
\end{sidewaystable}

The number of quanta is again found using Eqs.(\ref{consm1}) and (\ref{consm2}) using the same value for the normalization factor, $\gamma$,  given in Eq.(\ref{norm}). We  take  $m_H=130$ GeV here and in the rest of this section. We plot the number of quanta as a function of the classicalon mass in Fig.~\ref{ntopp}   for the two different choices, $M_*=500$ GeV and $M_*= 1$ TeV for both positive and negative $\kappa$. The  curves are again almost linear as in the previous case of goldstone classicalization. We also show for comparison the $N_*$ vs $M$ curve for a black hole in $n=3$ extra dimensions with $M_{pl}=400$ GeV. Once again,  for comparison with the black hole multiplicity, it must be kept in mind that  the final decay multiplicity, in the classicalon case, is bigger than $N_*$, the number of  Higgs bosons, as the Higgs bosons decay further to leptons and jets.
\begin{figure}[t]
\centering
\hspace{-0 in}
\begin{tabular}{cc}

\includegraphics[width=0.7\columnwidth]{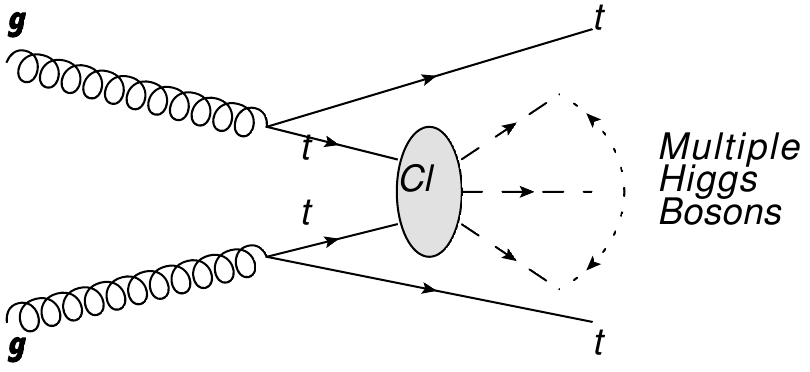}
\end{tabular}
\caption{Production of a classicalon by the top fusion process in the model with the Higgs as the classicalizer..}
\label{topf}
\end{figure}

We will consider the possibility of producing Higgsions in the top fusion process $gg\to  (t\bar{t} \to Cl)t\bar{t}$ (see Fig.~\ref{topf}). To find the cross-section  for classicalon production form top fusion  we introduce a dimension-5 $tthh$ operator in CALCHEP~Ê\cite{Pukhov:2004ca}, ($\bar{t}t H^\dagger H/\Lambda$) and find the cross-section for the top fusion process $pp \to gg \to t\bar{t}hh$. At high energies the $tt\to hh$ cross-section due to this operator is a constant as a function of the $t\bar{t}$-energy; this is also true for production of $N_*$-particle classicalons ($tt \to Cl$) the cross-section in this case being fixed at $\pi r_*^2 (M_{N_*})$. Thus we  find the cross-section of  the top fusion process $pp\to gg\to t \bar{t}hh$ (taking into account only the contribution of the $tthh$ operator and not other SM processes) and rescale this cross-section by the ratio of the $t \bar{t} \to Cl$ cross-section to the $tt \to hh$ cross-section to obtain the $pp\to gg\to t\bar{t} +Cl$ cross-section.  Finally we would have to multiply by a factor of 1/2 as the $tt \to Cl$ process would take place only if both the tops are right handed whereas for the $tt \to hh$ process to take place the tops need to have opposite chiralities. Note that we are assuming that the total cross-section can be factorized into a hard part and ($t\bar{t} \to Cl$) and a `top parton density function (PDF)' and this is not expected to be accurate unless the partonic center of mass energy $\sqrt{\hat{s}} \gg m_t$. For this reason our cross-section estimates would be approximate.

 We show the results for the cross-sections for  $M_*=500$ GeV and $M_*= 1$ TeV   in Table 2. We also give branching ratios and  cross-sections for the $l+\Emisst +jets$ channel (again requiring exactly, and not at least,  one lepton) where the lepton comes from a real or virtual $W$ boson emerging form either a Higgs or one of the final tops (decay channels with greater number of leptons have a much smaller branching fraction  in this case). Again, unlike black holes, missing energy must necessarily be present in this channel. The number of  jets is even larger here and a classicalon with $N_*=6$ would give rise to about  16 jets including the jets from the top decays, so that the background is again negligible~\cite{Lisanti:2011tm}. A similar estimate to the one done in the previous subsection tells us that for 16 jets at least  about $50\%$ of the theoretical cross-section  should survive after the lepton identification  and isolation requirements are taken into account. It is clear from Table 2 that  discovery would not be possible in the 7 TeV run of the LHC. For  $M_*=500$ GeV discovery should be possible with about 10 fb$^{-1}$ at 14 TeV LHC energy for both the $\kappa<0$ and $\kappa>0$ cases. Much higher integrated luminosities, about 100 fb$^{-1}$, would be required for $M_*=1$ TeV and $\kappa<0$ whereas  the $\kappa>0$ case would be out of reach even with high luminosities.

\section{Conclusions}

We have argued that classicalons must have analogs of  thermodynamic properties and we have carried out a model-independent statistical mechanical analysis of classicalons. By taking  the set of four momenta of the incoming (outgoing) particles that form a classicalon (that a classicalon decays to) as a  microstate of the classicalon, we count the number of such microstates imposing  only the condition of energy-momentum conservation and the condition that the incoming wave-packets should be able to localize their energy inside the classicalon radius, $r_*$. We  find that the particles a classicalon decays to will have a Planck distribution with an effective temperature $T \sim 1/r_*$ in the case of  a massless classicalizer field.  The  final thermodynamic relations obeyed by a classicalon are different  from those obeyed by  blackbody radiation in spite of the fact that both have the same distribution function. This is  because incoming/outgoing wave-packets in a classicalon formation/decay process have a different density of states than the particles in blackbody radiation. Our results confirm the expectations of Ref.~\cite{Dvali:2011th} and we find the entropy scales like, $S\sim N_* \sim M r_*$, when the  classicalizer field is massless.  This implies that classicalon decays to a few particles should be combinatorially suppressed by a factor $e^{-S} \sim e^{-N_*}$. For the specific case of a black hole, the classicalon radius is proportional to its mass, and the well known proportionality of the black hole entropy to its area follows from the general scaling of the classicalon entropy.

We use our results, in particular the computation of the number of classicalon decay products,  $N_*$,  to make LHC predictions. For computing the rate of production, we use the fact that classicalons are expected to be produced with a geometric cross-section, $\pi r_*^2$. The important difference from black hole production is that even at energies higher  than the classicalization scale, other SM processes involving particles without a strong classicalizing interaction go on unaffected with a larger cross-section than classicalon production. In the models we consider, light quarks have no direct classicalizing interactions and, as a result, the classicalon production cross-sections are much smaller than black hole production cross-sections at the same energy. On the other hand, we find the multiplicity of final decay products of the classicalons to be  larger  than the decay multiplicity of extra-dimensional black holes, in the cases we consider. 

The first model we look at is a model where longitudinal $WW$ scattering is unitarized in the absence of a Higgs by classicalization of longitudinal $W$s and $Z$s. The classicalon in this model decays to multiple $W$s and $Z$s which lead to signals in various channels like $l  +\Emisst + jets$, $l^+l^+  +\Emisst +jets$, $3l +\Emisst +jets$ and   $3l^+ +\Emisst  +jets$ where the number of partonic jets is typically larger than ten. Our results for the different channels are well summarized in Fig.~\ref{bars}. We find that, for this model, discovery would be imminent in the $l+ jets$ channel in the present 7 TeV run of the LHC, if the classicalization scale is as low as $M_* = v=246$ GeV and that we would have to wait for  about 10 fb$^{-1}$ integrated lumiosity at 14 TeV,  if the scale is higher, around $M_* =600$ GeV. The maximum classicalization scale that can be probed with 100 fb$^{-1}$ data at 14 TeV is about $M_*=1$ TeV. 

For the model to address the hierarchy problem with the Higgs itself as the classicalizer, we consider the minimal case where only the right handed top has a classicalizing interaction. The classicalon radius in this case depends on the sign of the non-renormalizable coupling $\kappa$. We explore the prospect of discovery of such classicalons in the top fusion process $gg\to  (t\bar{t} \to Cl)t\bar{t}$  by looking at the $l +\Emisst + jets$ final state where the number of partonic jets is very high (at least 15). We find that  for  $M_*=500$ GeV, discovery should be possible with about 10 fb$^{-1}$ at 14 TeV LHC for both the $\kappa<0$ and $\kappa>0$ cases. For $M_*=1$ TeV and $\kappa<0$ much higher integrated luminosities, about 100 fb$^{-1}$, would be required whereas  the $\kappa>0$ case would be out of reach even with high luminosities if $M_*=1$ TeV. 

Thus, we have shown that classicalon decays can produce remarkable multi-W/Z or multi-Higgs signatures at the LHC.  Our encouraging results on the discovery prospects of classicalization suggest that a more rigorous experimental study including event generation, QCD and detector effects   should be undertaken  in the future.

\noindent
{\bf Acknowledgments} 

\noindent This project would not be possible without crucial  inputs from  Gia Dvali and Gian Giudice. We would also like to thank  I. Florakis, C. Gomez, A. Khmelnitsky, A. Mertens, A. Vartak, A. Vikman and especially D. Pappadopulo and J.D. Wells for  valuable conversations. This work was supported in part by the European Commission under the contract ERC advanced grant 226371 MassTeV and the contract PITN-GA-2009-237920 UNILHC.
\section*{APPENDIX A: Transverse length of wave-packets forming a classicalon}

\begin{figure}
\begin{center}
\includegraphics[width=0.8\columnwidth]{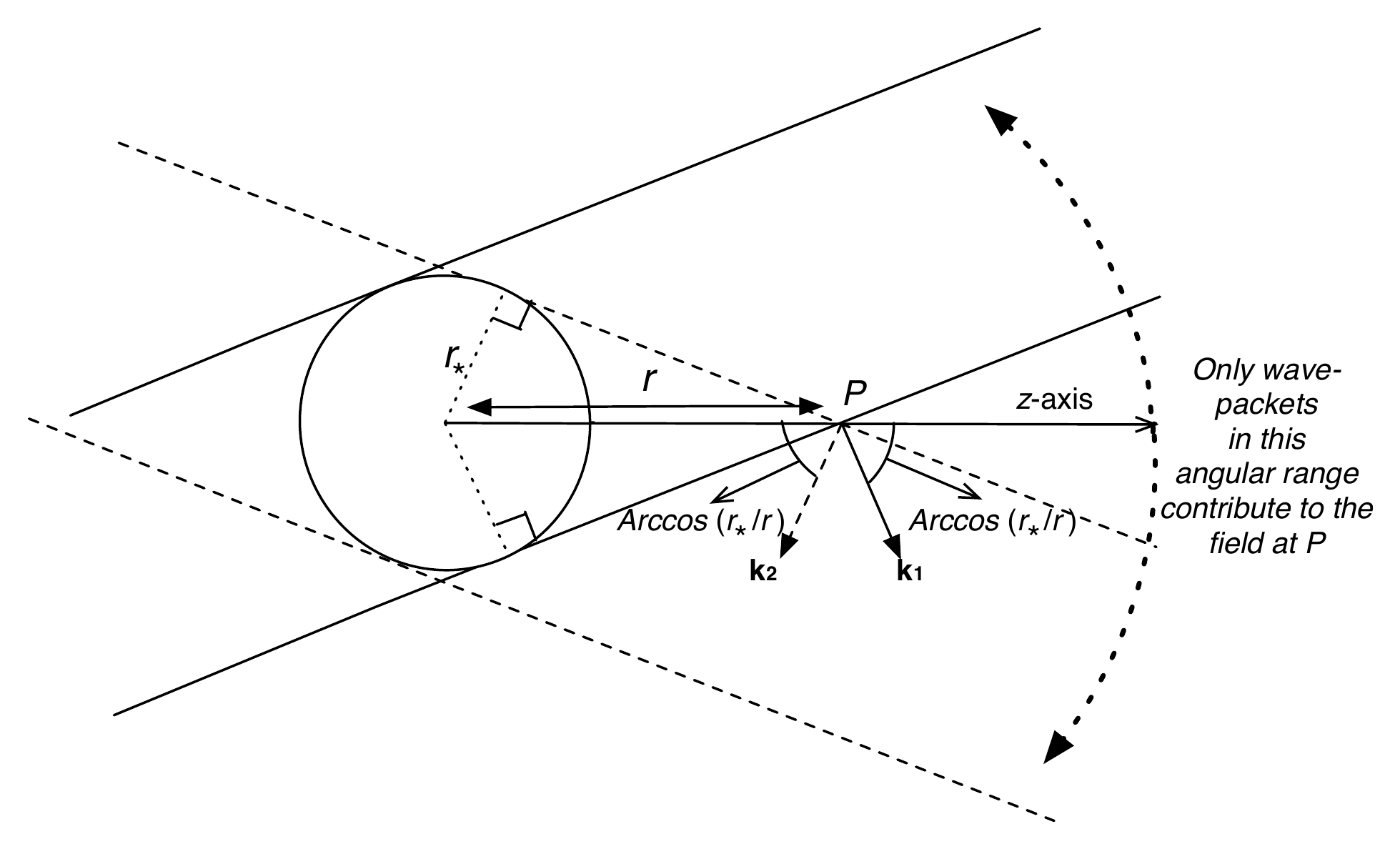}
\end{center}
\caption{ Here we show that at $t=0$ when all the wave-packets reach the origin, only wave-packets with direction of momentum in the $\theta$ range $-\cos^{-1}(r_*/r)<\theta<\cos^{-1}(r_*/r)$ contribute to the field  at a point $P$ on the $z$-axis at a distance $r$ from the origin. Here $\vec{k}_1$ and $\vec{k}_2$ are the momentum vectors of the two wave-packets shown}
\label{geom}
\end{figure}

We want to show in this Appendix that while the transverse length of  wave-packets forming a classicalon can be much larger than $2r_*$, it cannot be larger than $\sqrt{N} r_*$, where $N$ is the number of incoming wave-packets. We will show, first of all, that when all the incoming wave-packets reach the origin at $t=0$  (see Fig.~\ref{tlz}(right)), because of the transverse length being larger than $2 r_*$, there is a field  $\phi$ outside the classicalon radius $r_*$ but it drops off as $\phi \sim 1/r$. To prove this,   let us think for the moment,  although as we will soon see this cannot be the case, that the wave-packets are infinitely extended in the transverse direction.  If the number of these wave-packets is very large we can approximate the summation in the superposition of these wave-packets by an integral over a spherically symmetric distribution of these wave-packets with the direction of the  momenta $\vec{k}$ varying continuously.  Let $\theta$ be the angle the momentum of a particular wave-packet makes with the $z$-axis (see Fig.~\ref{geom}). For a point $P$ on the $z$-axis outside the sphere, at a distance $r$ from the origin, only wave-packets with direction of momentum in a certain $\theta$ range, $-\cos^{-1}(r_*/r)<\theta<\cos^{-1}(r_*/r)$, contribute  to the field $\phi$ (see Fig.~\ref{geom}) if $r>r_*$. On the other hand, for a point inside the classicalon, there are contributions from all the wave-packets without a restriction on  $\theta$. The total contribution to the field $\phi$ at  $P$, at a distance $r$ from the origin,  from  wave-packets with  energy $|\vec{k}|$ is,
\begin{figure}
\begin{center}
\includegraphics[width=0.7\columnwidth]{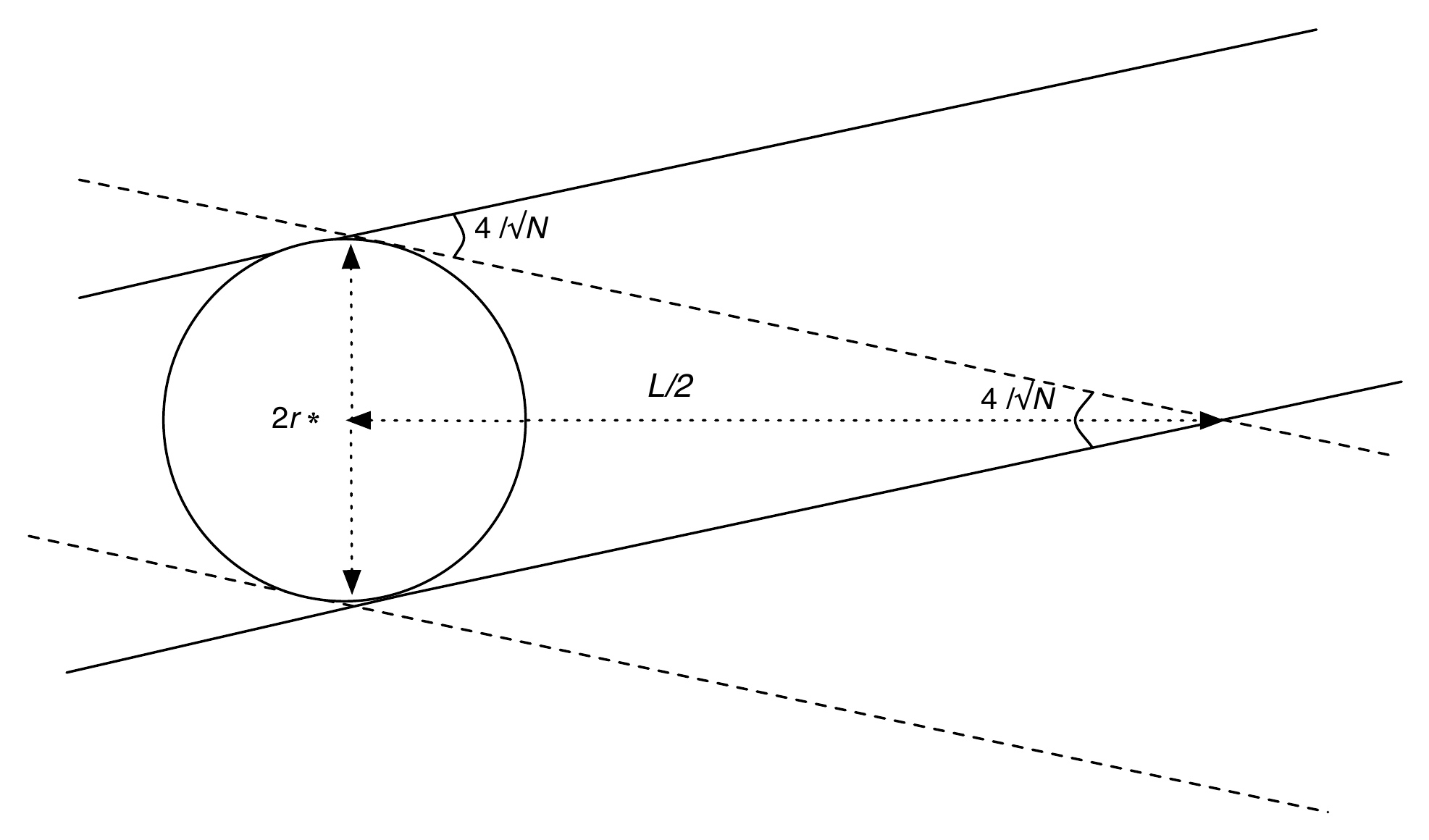}
\end{center}
\caption{ The distance $L/2$ at which two neighboring wave-packets stop overlapping. }
\label{over}
\end{figure}
\bea
\phi(r) &\sim& \int_{-1}^{1} \sin k (r \cos \theta+ r_*) d(\cos  \theta)  \int_0^{2 \pi} d \phi~~~~{\rm for~} r<r_*,\nonumber\\
\phi(r) &\sim& \int_{-r_*/r}^{r_*/r} \sin k( r \cos \theta+ r_*) d(\cos  \theta)  \int_0^{2 \pi} d \phi \sim \frac{1}{r}~~~~{\rm for~} r> r_*,
\label{obrB}
\eea
where we have used the functional form in Eq.~\ref{loung} and substituted $kl= \vec{k}.\vec{r}=k r \cos \theta$, $\vec{r}$ being the position vector of the point $P$. The exact form of the function inside the radius $r_*$ is not important as it would change anyway in the presence of a classicalizing interaction. What is important is the $\phi \sim 1/r$ drop off  outside the radius $r_*$ which shows that most of the the energy does get localized  inside $r_*$ for these wave-packets (note that the energy density goes as $(\partial \phi)^2\sim 1/r^4$).  We can also obtain the normalization $Q$ of the field, in $\phi \sim Q/r$, at the parametric level. For this note that our wavepackets must have the usual normalization $1/ \sqrt{\omega N r_*^3}$, $N r_*^3$ being the total volume of the wave-packets. Keeping in mind that the number of wave-packets giving a contribution in Eq.(\ref{obrB}) in the interval between $(\theta, \phi)$ and $(\theta+ d\theta, \phi+d\phi)$ is $(N/4\pi) d (\cos \theta) d \phi$ we  get,
\beq
\phi(r)=  \frac{N}{4 \pi}\frac{1}{\sqrt{\omega N r_*^3}}\int_{-r_*/r}^{r_*/r} \sin k( r \cos \theta+ r_*) d (\cos \theta) \int_0^{2 \pi} d \phi \sim \frac{\sqrt{N}}{r}~~~~{\rm for~} r > r_*.
\eeq
where we have used  the typical value,  $\omega=k=1/r_*$. The numerator  $\sqrt{N} = \sqrt{M r_*}$ is actually the correct charge in any classicalizing theory~\cite{Dvali:2010jz}. For instance in the special case of black holes it correctly reduces to the mass $M$ of the black hole.

 As we argued in Section~\ref{geomm}, in order that they always overlap, our wave-packets must have a finite transverse length. Let us calculate this length for $N$ incoming/outgoing particles. For $N$ particles at a radius $L/2$ from the origin, each particle can be thought to occupy an area $\pi L^2/N$ where no other particle is present. Assuming this area occupied by the particle to be circular we find that  on an average the angle between the momenta of two neighboring particles would be $(2 L/\sqrt{N})/(L/2)=4/\sqrt{N}$. As one can see from Fig.~\ref{over} this would mean that two neighboring wave-packets would stop overlapping at a distance $L/2$ given by,
\beq
\frac{L}{2} \times \frac{4}{\sqrt{N}}=2 r_*\Rightarrow L= \sqrt{N} r_*.
\label{overlapB}
\eeq
\section*{APPENDIX B: Derivation of the Bose-Einstein distribution function }

In this Appendix we  review the standard textbook derivation of the Bose-Einstein distribution function (for a more detailed treatment see, for instance, Ref.~\cite{tolman}). We want to find the distribution function $N_\omega$ that maximizes $\Omega(M)$   while respecting the energy conservation constraint,
\beq
\sum_\omega N _\omega g_\omega \omega~ d \omega=M.
\label{con2B}
\eeq
 Remember that $N_\omega$ is the number of particles in the energy state with energy  $\omega$, and $g_\omega$ is the degeneracy of this energy state.   Let us represent an arbitrary configuration of a particular energy level as, $ \times \times | \times |...\times$ where the crosses represent the indistinguishable $\phi$ quanta and the space between two bars is a quantum state. Thus we should have $N_\omega$ crosses and $g_\omega -1$ bars and the number of ways of arranging these  crosses and  bars would give us the number of ways of arranging the particles in a particular energy level. Considering all energy levels, this leads to the well known expression,
\beq
\Omega(M)=\Pi_\omega \frac{(N_\omega
+g_\omega)!}{N_\omega! g_\omega!}
\eeq
where we have approximated $(N_\omega
+g_\omega-1)! \approx (N_\omega
+g_\omega)!$ and $(g_\omega-1)!\approx g_\omega!$. We want to  maximize the entropy,
$S=\log (\Omega(M))$, respecting the constraints in Eq.(\ref{con2B}). We must have,
\bea
d S=\sum \log \frac{N_\omega + g_\omega}{g_\omega} dN_\omega=0 
\label{snm1} \\
dM =\sum \omega   dN_\omega=0 \label{beta}.
\label{snm2}
\eea
where we have used Stirling's approximation, $\log N!=N \log N- N$. Now we maximize $S$ by using the method of Lagrange multipliers,
\bea
dS-\beta dM &=&0 \\
\Rightarrow \log\left( 1+ \frac{g_\omega}{N_\omega}\right)- \beta \omega &=&0,
\label{defB}
\eea
where we have used Eqs.(\ref{snm1}) and (\ref{snm2}) and  $\beta$ is the Lagrange multiplier.  This leads to the Bose-Einstein distribution,
\beq
N_\omega = \frac{g_\omega d \omega}{e^{\beta \omega}-1}.
\eeq
Note that in our case there is no constraint on the total number of particles. Such a constraint would have led to the presence of a chemical potential which is zero in our case.

\section*{APPENDIX C: Branching ratios in goldstone classicalization}

In this Appendix we will provide formulae for the branching ratios of a classicalon to final states with varying number of leptons in the goldstone classicalization model.  In the expressions below, $w_l$ is the branching ratio of  a $W$ to leptons ($e$, $\mu$ and leptonically decaying $\tau$s), $w_j$ is the branching ratio of a $W$ to jets (including hadronically decaying $\tau$s ), $z_l$ is the branching ratio of a $Z$ to two leptons and  $z_j$ is the branching ratio of a $Z$ to two  jets. In general we include invisible decays of the $Z$ in $z_j$. This gives $w_l= 0.25$, $w_j= 0.75$, $z_l= 0.91$ and $z_j = 0.07$. To compute branching ratios to final states with maximum possible number of jets in association with a given number of leptons, we  do not include  $W$s decaying to hadronically decaying $\tau$s  in $w_j$ and  invisibly decaying  $Z$s in $z_j$, which changes the values of $w_j$ and $z_j$ above to $w_j=0.68$ and $z_j= 0.71$. The branching ratios depend on the electric charge, $Q$, of the classicalon. Let us first consider neutral classicalons, i.e. the $Q=0$ case.

\noindent
\textbf{Classicalons with $Q=0$}

In a general configuration for a neutral classicalon there are $k$ $W^+W^-$ pairs and $(N_*-2k)$ $Z$-bosons where $0\leq k \leq [N_*/2] $,  $ [N_*/2]$ being the largest integer smaller than $N_*/2$. As explained in Section~\ref{brs},Ê the probability of having such a configuration is given by,
\bea
P'_k&=&\frac{N_* !}{(N_*-k)! (k!)^2},\nonumber\\
P_k&=&\frac{P'_k}{ \sum_{k=0}^{n} P'_k}.
\eea
In order to obtain the single lepton final state one of the $W$s needs to decay leptonically which gives the following branching ratio for a $N_*$-particle classicalon,
\beq
BR(Cl \to l)=\sum_{k=1}^{ [N_*/2]}  \binom{2k}{1} w_l w_j^{2k-1} z_j^{N_*- 2k} P_k.
\eeq
Similarly to obtain two(three) positive leptons, two (three) $W^+$s need to decay leptonically which leads to the expressions,
\bea
BR(Cl \to l^+l^+ )=\sum_{k=2}^{ [N_*/2]}  \binom{k}{2} w_l^2 w_j^{2k-2} z_j^{N_*- 2k}P_k, \\
BR(Cl \to 3l^+ )=\sum_{k=3}^{ [N_*/2]} \binom{k}{3} w_l^3 w_j^{2k-3} z_j^{N_*- 2k}P_k. 
\eea
For the branching ratio to 3 leptons either 3 $W$s need to decay leptonically or 2 $W$s and a $Z$ need to decay leptonically which gives us two terms in the branching ratio of a classicalon to 3 leptons, 
\beq
BR(Cl \to 3l)=\sum_{k=2}^{ [N_*/2]} \binom{2k}{3} w_l^3 w_j^{2k-3} z_j^{N_*- 2k}P_k+\sum_{k=1}^{ [N_*/2]}  \binom{2k}{1} w_l w_j^{2k-1}\binom{N_*- 2k}{1} z_l z_j^{N_*- 2k-1}P_k. 
\eeq
Now let us generalize this to a classicalon decay to an arbitrary number of leptons, $n_l$ where $0\leq n_l \leq 2 N_*$.  For $n_l=2p$, an even number, we can get $n_l$ leptons from the decay of  an even number, $2q$, of $W$ decays and $(p-q)$,  $Z$ decays. This gives us,
\beq
BR(Cl \to2p~l)=\sum_{q=0}^{p}\sum_{k=q}^{ [N_*/2]} \binom {2k}{2q} w_l^{2q} w_j^{2k-2q}\binom{N_*-2k}{p-q} z_l^{p-q} z_j^{N_*- 2k-p+q}P_k. 
\eeq
If $n_l=(2p +1)$, is an odd number, we can get $n_l$ leptons from the decay of  an odd number, $(2q+1)$, of $W$ decays and $(p-q)$, $Z$ decays. This gives us,
\beq
BR(Cl \to(2p+1)l )=\sum_{q=0}^{p}\sum_{k=q}^{ [N_*/2]} \binom{2k}{2q+1} w_l^{2q+1} w_j^{2k-2q-1}\binom{N_*-2k}{p-q} z_l^{p-q} z_j^{N_*- 2k-p+q}P_k.
\eeq

\noindent
\textbf{Classicalons with $Q=+2$}

For classicalons with charge $Q=+2$, there are in general $(k+2)$ $W^+$ bosons, $k$ $W^-$ bosons and $(N_*-2k-2)$ $Z$-bosons, where $0\leq k \leq [(N_*-2)/2]$, $[(N_*-2)/2]$ being the largest integer smaller than $(N_*-2)/2$. Proceeding as in the previous case we obtain the expressions,
\bea
P'_k&=&\frac{N_* !}{(N_*-2k-2)! k!(k+2)!},\nonumber\\
P_k&=&\frac{P'_k}{ \sum_{k=0}^{[(N_*-2)/2]} P'_k},\\
BR(Cl \to l)&=&\sum_{k=0}^{[(N_*-2)/2]} \binom{2k+2}{1} w_l w_j^{2k+1} z_j^{N_*- 2k-2}P_k,\\
BR(Cl \to l^+l^+)&=&\sum_{k=0}^{[(N_*-2)/2]} \binom{k+2}{2} w_l^2 w_j^{2k} z_j^{N_*- 2k-2}P_k, \\
BR(Cl \to 3l^+)&=&\sum_{k=1}^{[(N_*-2)/2]}  \binom{k+2}{3} w_l^3 w_j^{2k-1} z_j^{N_*- 2k-2}P_k,
\eea
\bea
BR(Cl \to 3l)&=&\sum_{k=1}^{[(N_*-2)/2]} \binom {2k+2}{3} w_l^3 w_j^{2k-1} z_j^{N_*- 2k-2} P_k
\nonumber\\
&&+\sum_{k=0}^{[(N_*-2)/2]} \binom{2k+2}{1} w_l w_j^{2k+1}\binom{N_*- 2k-2}{1} z_l z_j^{N_*- 2k-3}P_k,\\
BR(Cl \to 2p~Ê l )&=&\sum_{q=0}^{p}\sum_{k=q-1}^{[(N_*-2)/2]} \binom {2k+2}{2q} w_l^{2q} w_j^{2k-2q+2}  \times \nonumber\\ &&~~~~~~~~~~~~~~\binom{N_*-2k-2}{p-q} z_l^{p-q} z_j^{N_*- 2k-2-p+q} P_k,\\
BR(Cl \to(2p+1) l )&=&\sum_{q=0}^{p}\sum_{k=q-1}^{[(N_*-2)/2]}  \binom{2k+2}{2q+1} w_l^{2q+1} w_j^{2k-2q+1}  \times \nonumber\\ &&~~~~~~~~~~~~~~
\binom{N_*-2k-2}{p-q} z_l^{p-q} z_j^{N_*- 2k-2-p+q}P_k.
\eea

\noindent
\textbf{Classicalons with $Q=-2$}

For classicalons with charge $Q=+2$, there are  in general $(k+2)$ $W^-$ bosons, $k$ $W^+$ bosons and $(N_*-2k-2)$ $Z$-bosons, where $0\leq k \leq [(N_*-2)/2]$, $[(N_*-2)/2]$ being the largest integer smaller than $(N_*-2)/2$. In this case we obtain,
\bea
P'_k&=&\frac{N_* !}{(N_*-2k-2)! k!(k+2)!},\nonumber\\
P_k&=&\frac{P'_k}{ \sum_{k=0}^{[(N_*-2)/2]} P'_k},\\
BR(Cl \to l)&=&\sum_{k=0}^{[(N_*-2)/2]} \binom{2k+2}{1} w_l w_j^{2k+1} z_j^{N_*- 2k-2}P_k,\\
BR(Cl \to l^+l^+)&=&\sum_{k=2}^{[(N_*-2)/2]} \binom{k}{2} w_l^2 w_j^{2k} z_j^{N_*- 2k-2}P_k, \\
BR(Cl \to 3l^+)&=&\sum_{k=3}^{[(N_*-2)/2]}  \binom{k}{3} w_l^3 w_j^{2k-1} z_j^{N_*- 2k-2}P_k,
\eea
\bea
BR(Cl \to 3l)&=&\sum_{k=1}^{[(N_*-2)/2]} \binom {2k+2}{3} w_l^3 w_j^{2k-1} z_j^{N_*- 2k-2} P_k
\nonumber\\
&&+\sum_{k=0}^{[(N_*-2)/2]} \binom{2k+2}{1} w_l w_j^{2k+1}\binom{N_*- 2k-2}{1} z_l z_j^{N_*- 2k-3}P_k,\\
BR(Cl \to 2p~Ê l )&=&\sum_{q=0}^{p}\sum_{k=q-1}^{[(N_*-2)/2]} \binom {2k+2}{2q} w_l^{2q} w_j^{2k-2q+2}  \times \nonumber\\ &&~~~~~~~~~~~~~~\binom{N_*-2k-2}{p-q} z_l^{p-q} z_j^{N_*- 2k-2-p+q} P_k,\\
BR(Cl \to(2p+1) l )&=&\sum_{q=0}^{p}\sum_{k=q-1}^{[(N_*-2)/2]}  \binom{2k+2}{2q+1} w_l^{2q+1} w_j^{2k-2q+1}  \times \nonumber\\ &&~~~~~~~~~~~~~~
\binom{N_*-2k-2}{p-q} z_l^{p-q} z_j^{N_*- 2k-2-p+q}P_k.
\eea


\begin{thebibliography}{5}
\bibitem{Dvali:2010jz}
  G.~Dvali, G.~F.~Giudice, C.~Gomez, A.~Kehagias,
  ``UV-Completion by Classicalization,''
  [arXiv:1010.1415 [hep-ph]].
\bibitem{atlas0}
 See, for instance, the talk by W. Murray in the 2011 Europhysics Conference On High Energy Physics, Grenoble  (http://is.gd/UJl43n).
\bibitem{Dvali:2010bf}
  G.~Dvali, C.~Gomez,
  ``Self-Completeness of Einstein Gravity", [arXiv:1005.3497 [hep-th]].
\bibitem{Dvali:2010ns}
  G.~Dvali, D.~Pirtskhalava,
  ``Dynamics of Unitarization by Classicalization,''
  Phys.\ Lett.\  {\bf B699}, 78-86 (2011).
  [arXiv:1011.0114 [hep-ph]].
\bibitem{Dvali:2011nj}
  G.~Dvali,  ``Classicalize or not to Classicalize?,''  [arXiv:1101.2661 [hep-th]].
\bibitem{Dvali:2011th}
  G.~Dvali, C.~Gomez, A.~Kehagias,
  ``Classicalization of Gravitons and Goldstones,''
   [arXiv:1103.5963 [hep-th]].
   \bibitem{Bajc:2011ey}
  B.~Bajc, A.~Momen, G.~Senjanovic,
  ``Classicalization via Path Integral,''  [arXiv:1102.3679 [hep-ph]].
  \bibitem{Akhoury:2011en}
  R.~Akhoury, S.~Mukohyama, R.~Saotome, ``No Classicalization Beyond Spherical Symmetry,''
   [arXiv:1109.3820 [hep-th]].
  \bibitem{Brouzakis:2011zs}
  N.~Brouzakis, J.~Rizos, N.~Tetradis,
  ``On the dynamics of classicalization,''  [arXiv:1109.6174 [hep-th]].
 \bibitem{greiner}
 See for eg. Pg. 153 of W.Greiner, L.Neise and H.Stocker â ``Thermodynamics and Statisti- cal Mechanics," Springer (1997).
\bibitem{Antoniadis:1998ig}
  I.~Antoniadis, N.~Arkani-Hamed, S.~Dimopoulos, G.~R.~Dvali,
  ``New dimensions at a millimeter to a Fermi and superstrings at a TeV,''
  Phys.\ Lett.\  {\bf B436}, 257-263 (1998).
  [hep-ph/9804398].
  \bibitem{Dvali:2001gx}
  G.~R.~Dvali, G.~Gabadadze, M.~Kolanovic, F.~Nitti,
  ``Scales of gravity,''
  Phys.\ Rev.\  {\bf D65}, 024031 (2002).
  [hep-th/0106058].
  \bibitem{Giddings:2001bu}
  S.~B.~Giddings, S.~D.~Thomas,
  ``High-energy colliders as black hole factories: The End of short distance physics,''
  Phys.\ Rev.\  {\bf D65}, 056010 (2002).
  [hep-ph/0106219].
\bibitem{Dimopoulos:2001hw}
  S.~Dimopoulos, G.~L.~Landsberg,
  ``Black holes at the LHC,''
  Phys.\ Rev.\ Lett.\  {\bf 87}, 161602 (2001).
  [hep-ph/0106295].

\bibitem{Giudice:2001ce}
  G.~F.~Giudice, R.~Rattazzi, J.~D.~Wells,
  ``Transplanckian collisions at the LHC and beyond,''
  Nucl.\ Phys.\  {\bf B630}, 293-325 (2002).
  [hep-ph/0112161].
  \bibitem{Stirling:2011mf}
  W.~J.~Stirling, E.~Vryonidou, J.~D.~Wells,
  ``Eikonal regime of gravity-induced scattering at higher energy proton colliders,''
  Eur.\ Phys.\ J.\  {\bf C71}, 1642 (2011).
  [arXiv:1102.3844 [hep-ph]].
  \bibitem{Meade:2007sz}
  P.~Meade, L.~Randall,
  ``Black Holes and Quantum Gravity at the LHC,''
  JHEP {\bf 0805}, 003 (2008).
  [arXiv:0708.3017 [hep-ph]].
\bibitem{Dvali:2011nh}
  G.~Dvali, C.~Gomez, S.~Mukhanov,
  ``Black Hole Masses are Quantized,'' [arXiv:1106.5894 [hep-ph]].
  \bibitem{Lee:1977eg}
  B.~W.~Lee, C.~Quigg, H.~B.~Thacker,
  ``Weak Interactions at Very High-Energies: The Role of the Higgs Boson Mass,''
  Phys.\ Rev.\  {\bf D16}, 1519 (1977).

  \bibitem{Barbieri:2004qk}
  R.~Barbieri, A.~Pomarol, R.~Rattazzi, A.~Strumia,
  ``Electroweak symmetry breaking after LEP-1 and LEP-2,''
  Nucl.\ Phys.\  {\bf B703}, 127-146 (2004).
  [hep-ph/0405040].
  \bibitem{Dutta:2007st}
  S.~Dutta, K.~Hagiwara, Q.~-S.~Yan, K.~Yoshida,
  ``Constraints on the electroweak chiral Lagrangian from the precision data,''
  Nucl.\ Phys.\  {\bf B790}, 111-137 (2008).
  [arXiv:0705.2277 [hep-ph]].
  \bibitem{BHpas}
  CMS Public Analysis Summary, ``Search for Black Holes in $pp$ Collisions at $\sqrt{s}=7$ TeV
 with 1 fb$^{-1}$ Data Set", CMS PAS EXO-11-071.
\bibitem{Nakamura:2010zzi}
  K.~Nakamura {\it et al.} [ Particle Data Group Collaboration ],
``Review of particle physics,''
  J.\ Phys.\ G {\bf G37}, 075021 (2010).
  
\bibitem{Dawson:1984gx}
  S.~Dawson,
  ``The Effective W Approximation,''
  Nucl.\ Phys.\  {\bf B249}, 42-60 (1985).
\bibitem{Martin:2009iq}
  A.~D.~Martin, W.~J.~Stirling, R.~S.~Thorne, G.~Watt,
  ``Parton distributions for the LHC,''
  Eur.\ Phys.\ J.\  {\bf C63}, 189-285 (2009).
  [arXiv:0901.0002 [hep-ph]].
\bibitem{Evans:2009ga}
  J.~A.~Evans, M.~A.~Luty,
  ``Strong Electroweak Symmetry Breaking and Spin 0 Resonances,''
  Phys.\ Rev.\ Lett.\  {\bf 103}, 101801 (2009).
  [arXiv:0904.2182 [hep-ph]].
\bibitem{Mukhopadhyaya:2010qf}
  B.~Mukhopadhyaya, S.~Mukhopadhyay,
  ``Same-sign trileptons and four-leptons as signatures of new physics at the CERN Large Hadron Collider,'' Phys.\ Rev.\  {\bf D82}, 031501 (2010).
  [arXiv:1005.3051 [hep-ph]].
\bibitem{Lisanti:2011tm}
  M.~Lisanti, P.~Schuster, M.~Strassler, N.~Toro, ``Study of LHC Searches for a Lepton and Many Jets,''   [arXiv:1107.5055 [hep-ph]].  
  \bibitem{Contino:2008hi}
  R.~Contino, G.~Servant,
  ``Discovering the top partners at the LHC using same-sign dilepton final states,''
  JHEP {\bf 0806}, 026 (2008).
  [arXiv:0801.1679 [hep-ph]].
\bibitem{Dissertori:2010ug}
  G.~Dissertori, E.~Furlan, F.~Moortgat, P.~Nef,
  ``Discovery potential of top-partners in a realistic composite Higgs model with early LHC data,''
  JHEP {\bf 1009}, 019 (2010).
  [arXiv:1005.4414 [hep-ph]].
\bibitem{tdr}
  G.~Aad {\it et al.} [ The ATLAS Collaboration ],
  ``Expected Performance of the ATLAS Experiment - Detector, Trigger and Physics,'' [arXiv:0901.0512 [hep-ex]].

\bibitem{Pukhov:2004ca}
  A.~Pukhov,
  ``CalcHEP 2.3: MSSM, structure functions, event generation, batchs, and
  generation of matrix elements for other packages,''
  arXiv:hep-ph/0412191.
  \bibitem{tolman}
 See for instance Sections 87- 89 of  R.C.~Tolman,
  ``The Principles of Statistical Mechanics,''
Dover (1979).

\end{thebibliography}
\end{document}